\documentclass[12pt,a4paper]{article}
\usepackage[dvipdfmx]{graphicx,color}
\usepackage{bm}
\usepackage{amsmath,amssymb}
\usepackage{booktabs}
\makeatletter
\def\mbf#1{\mbox{\boldmath ${#1}$}}
\def\Slash{\mathpalette\@Slash}
\def\@Slash#1#2{{\ooalign{\hfil$#1/$\hfil\crcr$#1{#2}}}}
\makeatother
\begin{document}

\title{Leptogenesis in $E_6 \times U(1)_A$ SUSY GUT model}
\author{
\centerline{
Takuya~Ishihara$^{1}$\footnote{E-mail address: ishihara@eken.phys.nagoya-u.ac.jp}
,~
Nobuhiro~Maekawa$^{1,2}$\footnote{E-mail address: maekawa@eken.phys.nagoya-u.ac.jp}
,~ } 
\\*[5pt]
\centerline{
Mao~Takegawa$^{1}$%\footnote{E-mail address: ??@eken.phys.nagoya-u.ac.jp}
~and 
Masato~Yamanaka$^{2}$\footnote{E-mail address: yamanaka@eken.phys.nagoya-u.ac.jp}}
\\*[25pt]
\centerline{
\begin{minipage}{\linewidth}
\begin{center}
$^1${\it \normalsize Department of Physics, Nagoya University, Nagoya 464-8602, Japan }  \\*[10pt]
$^2${\it \normalsize Kobayashi Maskawa Institute, Nagoya University, Nagoya 464-8602, Japan }  \\*[10pt]\end{center}
\end{minipage}}
\\*[50pt]}
\date{}
\maketitle
\begin{abstract}
We study the thermal leptogenesis in the $E_6\times U(1)_A$ SUSY GUT model in which
realistic masses and mixings of quarks and leptons can be realized.
We show that the sufficient baryon number can be produced by the leptogenesis
in the model, in which the mass parameter of the lightest right-handed neutrino 
is predicted to be smaller than $10^8$ GeV. 
The essential point is that the mass 
of the lightest right-handed neutrino can be enhanced in the model because
it has a lot of mass terms whose mass parameters are predicted to be the same order of magnitude
which is smaller than $10^8$ GeV. 
We show that O(10) enhancement for the lightest right-handed neutrino mass is 
sufficient for the observed baryon asymmetry.
Note that such mass enhancements 
do not change 
the predictions of neutrino masses and mixings at the low energy scale
in the $E_6$ model which has six right-handed neutrinos.
In the calculation, we include the effects of supersymmetry and flavor in
final states of the right-handed neutrino decay. 
We show that the effect of supersymmetry is quite important even in the 
strong washout regime when the effect of flavor is included. This is because 
the washout effects on the asymmetries both of the muon and the electron 
become weaker than that of the tau asymmetry. 
\end{abstract}
 
\section{Introduction}
Supersymmetric (SUSY) grand unified theory (GUT)\cite{GUT} is one of the most promising candidates as the extended model of the standard model (SM). 
This is because the SUSY GUT realizes two kinds of unifications, unification of the gauge
 interactions and unification of the matters in the SM and for both unifications, there
 are supports from experiments. Three gauge couplings in the SM meets at a scale,
 which is called the GUT scale $\Lambda_G\sim 2\times 10^{16}$ GeV. Moreover, 
the various hierarchies of quark and lepton masses and mixings can be naturally understood
in $SU(5)$ unification if we assume that the $\bf 10$ fields of $SU(5)$ induce stronger
hierarchy in Yukawa couplings than the $\bf\bar 5$ fields of $SU(5)$.
One of the most important advantages of the $E_6$ unification\cite{E6GUT} is that the above assumption
can be naturally derived\cite{E6}. As the result of this important feature of the $E_6$ 
unification, we can build an $E_6$ GUT in which all three generation of quarks and leptons
can be unified into a single multiplet(or two multiplets) by introducing family symmetry
$SU(3)_F$(or $SU(2)_F$) and the realistic quark and lepton masses and mixings can be
realized after breaking the family and GUT symmetries\cite{E6F}.

However, it is well-known that SUSY GUTs are suffering from the doublet-triplet splitting
problem\cite{DT}. The doublet Higgs must have the weak scale mass to obtain the weak scale, while
the triplet (colored) Higgs which belongs to the same multiplet as the doublet Higgs 
in the GUT
must have the GUT scale mass to stabilize the nucleon. 
Fortunately, if the anomalous $U(1)_A$
gauge symmetry\cite{Ano} is introduced, the problem can be solved under a natural assumption
that all the interactions are introduced with $O(1)$ coefficients\cite{SO10, E6Higgs, gauge}. 
Because of this natural assumption, the coefficients of the terms and the vacuum 
expectation values (VEVs) of the GUT Higgs can be determined only by the symmetry of the
theory. The coefficients of the interaction $XYZ$ are determined\cite{FN, Ibanez:1994ig} 
except the $O(1)$coefficients by the total anomalous $U(1)_A$ charge $x+y+z$ as
\begin{equation}
\begin{array}{cl}
\lambda^{x+y+z}XYZ & (x+y+z\geq 0) \\
0 & (x+y+z<0),
\label{SUSYzero}
\end{array}
\end{equation}
where $x$, $y$, and $z$ are the $U(1)_A$ charges of the fields $X$, $Y$, and $Z$, 
respectively. Throughout this paper, we denote all the fields with uppercase letters and their anomalous $U(1)_A$ 
charges with the corresponding lowercase letters if there is no special comment.  Here $\lambda$ is the ratio 
of the Fayet-Illiopoulos parameter $\xi$ to
the cutoff $\Lambda$, and in this paper we take $\lambda\sim 0.22$ as a typical value. 
Under the natural assumption, we can obtain the realistic Yukawa couplings in $E_6$
GUT\cite{E6} (or in $SO(10)$ GUT\cite{SO10} which has similar structure as $E_6$ GUT).
The VEVs of the operators $O$ is also determined\cite{E6} by their total anomalous $U(1)_A$ charges 
$o$ as
\begin{equation}
\langle O\rangle=\left\{\begin{array}{ll}
                                0 & (o>0) \\
                                \lambda^{-o} & (o\leq 0)
                        \end{array}\right..
\label{VEV}
\end{equation}
In this paper, we often use a unit in which the cutoff $\Lambda$ is taken to be 1.
Because of the natural assumption, all the mass spectrum of superheavy particles and
the VEVs of GUT Higgs are determined only by the symmetry of the theory. Therefore,
we can calculate the running gauge couplings once we fix the symmetry of the theory.
Interestingly, this natural scenario gives a novel explanation\cite{gauge}
for the experimental support for the unification of three gauge interactions in the SM. 
The new explanation requires that the cutoff scale must be taken to be around the usual
 GUT scale $\Lambda_{G}$\cite{gauge}. 

If this natural $E_6$ GUT describes our world, it must be consistent with the cosmology.
The dark matter can be the lightest supersymmetric particle.
In this paper, we discuss the leptogenesis\cite{Fukugita:1986hr} in this scenario.
One of the important things in $E_6$ unification for the leptogenesis is that the 
fundamental representation
\textbf{27}, which is decomposed in the $E_6\supset SO(10)\times U(1)_{V'}$ notation 
(and in the [$SO(10)\supset SU(5)\times U(1)_V$] notation) as 
\begin{equation}
\bm{27}=\bm{16}_1[\bm{10}_{1}+\bar{\bm{5}}_{-3}+\bm{1}_{5}]+
\bm{10}_{-2}[\bm{5}_{-2}+\bm{\bar{5}'}_{2}]+\bm{1}'_4[\bm{1}'_0],
\end{equation}
 includes two singlets $S(\bm{1'})$ and $N_R^c(\bm{1})$ under the SM gauge group,
which can be the right-handed (RH)
neutrinos. If we introduce three \textbf{27} for three generation quarks and leptons,
we have six RH neutrinos. Basically, since 
the masses and Yukawa couplings of the 
RH neutrinos are determined by the symmetry, we can examine whether
the leptogenesis works well or not in this scenario.
Naively, the leptogenesis in this scenario does not work because the lightest RH
neutrino becomes lighter than $10^8$ GeV, i.e., this scenario looks not to satisfy
the Ibarra's upper bound\cite{Ibarra} for the lightest RH neutrino which is $10^{8-9}$ GeV. 
Actually, in a typical model, the (effective) $U(1)_A$ charges of 
$S_i$ and $N_{Ri}^c$ 
($i=1,2,3$) are fixed as $(\tilde s_1,\tilde s_2,\tilde s_3)=(6.5,5.5,3.5)$ and 
$(\tilde n_{R1}^c,\tilde n_{R2}^c,\tilde n_{R3}^c)=(6,5,3)$, and therefore, 
the mass of the lightest 
RH neutrino $S_1$ becomes $M_{S_1}\sim\lambda^{13}\Lambda\sim 5.7\times 10^7$ GeV\cite{E6}.
Yukawa couplings are also easily estimated because the sum of the (effective) $U(1)_A$
 charges
of the up-type Higgs $H_u$ and doublet-leptons $l_i$ become 
$(\tilde h_u+\tilde l_1,\tilde h_u+\tilde l_2,\tilde h_u+\tilde l_3)=(0,-0.5,-1)$. 
The Yukawa couplings among $l_i$, $S_1$ and $H_u$ become
$(\lambda^{6.5},\lambda^6,\lambda^{5.5})$. Then, we can estimate two important parameters
for the leptogenesis as
\begin{equation}
\begin{array}{l}
K\equiv \Gamma_D/H\sim 40 \\
\epsilon\equiv 
\frac{\Gamma(S_1\rightarrow l+H_u)-\Gamma(S_1\rightarrow \bar l+H_u^\dagger)} 
{\Gamma(S_1\rightarrow l+H_u)+\Gamma(S_1\rightarrow \bar l+H_u^\dagger)}\sim 
5\times 10^{-9},
\end{array}
\end{equation} 
where $\Gamma_D$ and $H$ are the decay width of $S_1$ and the Hubble parameter at $T=M_{S_1}$, respectively. 
(In this paper we denote the lepton doublet fields with lowercase letter $l$ in order to avoid the confusion with
lepton asymmetry $L$ in the following discussions. ) Since the sufficient production of Baryon number requires $K\sim 1$ and
$\epsilon\sim 10^{-7}$, this $K$ is too large,
and the $\epsilon$ is too small. The produced lepton number is calculated as
\begin{equation}
Y_L\equiv\frac{n_{L}}{s_0}\sim 10^{-13},
\end{equation}
which is about $O(1000)$ times smaller than the value $Y_L\sim 2.5\times 10^{-10}$ which
is required for the sufficient baryon number.
Here, $n_L$ and $s_0$ are the lepton number density and the entropy density today, and for simplicity, 
we neglect the SUSY contribution, which will be discussed later.

An important observation for leptogenesis in this scenario is that under fixed Yukawa 
couplings, $K\propto 1/M_{S_1}$ and $\epsilon\propto M_{S_1}$. Therefore, 
larger $M_{S_1}$ results in larger baryon number. This observation is critical because in this scenario,
the mass of $S_1$ tends to be larger than expected by the symmetry. There are two 
essential points in this scenario.
One of them
is that it has a plenty of terms which give mass to $S_1$. Each term gives the same order
of mass to $S_1$ as expected by the symmetry, and the real mass can increase because of
the large number of mass terms. The other point is that the predictions for the quark and
lepton masses and mixings does not change so much even if the mass of $S_1$ becomes
larger than expected by the symmetry. This is because the number of RH neutrino
 flavors becomes larger than three in $E_6$ unification. (In $SO(10)$ unification, it is not
 avoidable to change the predictions on neutrino sector if one of the RH 
 neutrino masses is taken to be larger than expected by the symmetry.)
 
 The question is how large enhancement of the mass is 
needed to obtain the sufficiently large baryon number. It is the main subject in this
paper to answer this question.

In section 2, we briefly review the $E_6$ GUT with anomalous $U(1)_A$ gauge symmetry.
And in section 3, we discuss the enhancement of the RH neutrino masses in
this scenario. In section 4, we would like to answer the above question.
First,  we calculate the enhancement factor required to obtain the sufficient baryon 
number with simple non-SUSY Boltzmann equations. Second, we discuss the effect of
lepton flavors. Third, the SUSY effect is considered. Finally, we calculate the 
enhancement factor, including both effects of lepton flavors and of SUSY.
We show that only about O(10) enhancement of the mass of $S_1$
 is sufficient for the observed baryon number.

\section{$E_6$ unification with anomalous $U(1)_A$ gauge symmetry}
We briefly review the $E_6\times U(1)_A$ GUT in this section\cite{E6}.
The typical quantum numbers of fields in $E_6\times U(1)_A$ GUT are
 shown in Table 1.
\begin{table}[tbp]
\begin{center}
\begin{tabular}{c|ccccccccccc}
 & $\Psi_1$ & $\Psi_2$ & $\Psi_3$ & $H$ & $\bar{H}$ & $C$ & $\bar{C}$ & $A$ \\
\hline
$E_6$ & ${\bf 27}$ & ${\bf 27}$ & ${\bf 27}$ & ${\bf 27}$ & ${\bf \overline{27}}$ &
${\bf 27}$ & ${\bf \overline{27}}$ & ${\bf 78}$ \\
$U(1)_A$ & $\frac{9}{2}$ & $\frac{7}{2}$ & $\frac{3}{2}$ & -3 & 1 & -4 & -1 & -1\\
\hline
\end{tabular}
\caption{Field contents of matters and GUT Higgs in a typical $E_6\times U(1)_A$ GUT\cite{E6}
and the charge assignment under 
$E_6\times U(1)_A$. Here,$\Psi_i$ $(i=1,2,3)$ are three generation quarks and leptons, 
the VEVs of $H$ and $\bar H$ break $E_6$ into $SO(10)$, 
the VEVs of $C$ and $\bar C$ break $SO(10)$ into $SU(5)$, and the VEV
of $A$ breaks $SU(5)$ into the standard model gauge group. The MSSM Higgs are included in $H$ and $C$. }
\label{tb:Field contents}
\end{center}
\end{table}
An interesting structure in $E_6$ unification is that three of six $\bar{\mbf{5}}$ of 
$SU(5)$ in three matter fields $\Psi_i$(\mbf{27}) become superheavy through the Yukawa interactions
\begin{equation}
(Y^H)_{ij}\Psi_i\Psi_jH+(Y^C)_{ij}\Psi_i\Psi_jC
\end{equation}
after developing the VEVs $\langle \bar HH\rangle\sim \lambda^{-h-\bar h}$ and
 $\langle \bar CC\rangle\sim \lambda^{-c-\bar c}$, which break $E_6$ into $SO(10)$ and
 $SO(10)$ into $SU(5)$, respectively. Here, the components of Yukawa matrices $Y^H$ and
 $Y^C$ are fixed by the total anomalous $U(1)_A$ charges of the corresponding terms $\Psi_i\Psi_jH$ and $\Psi_i\Psi_jC$, respectively.
Since the Yukawa couplings for $\Psi_3$ are larger than those for $\Psi_2$ and $\Psi_1$
because $\psi_3\gg \psi_1, \psi_2$, $\bar{\mbf{5}_3}$ and $\bar{\mbf{5}}_3^\prime$ become
superheavy, and therefore, three light modes $\bar{\mbf{5}}$ come from the $\Psi_1$ and $\Psi_2$.
This structure naturally explains why \mbf{10}s of $SU(5)$ induce stronger hierarchy than
$\bar{\mbf{5}}$s of $SU(5)$, which is important to obtain realistic hierarchies of quark
and lepton masses and mixings.

\begin{table}[p!]
\begin{center}
\caption{GUT scale $\Lambda_G$, 
Majorana masses of RH neutrinos $M_{\alpha} \, 
(\alpha = 1, 2, ..., 6)$, and each component of neutrino Yukawa 
$Y_{\nu}$ in the $E_6 \times U(1)_A$ GUT model with $\lambda=0.22$.}
\vspace{2mm}
\begin{tabular}{llllll}
\hline
Parameter
& value
& comment
\\ \hline \hline
$\Lambda_G$
& $2.000 \times 10^{16} \, \text{GeV}$
& GUT scale and the cutoff scale
\\[0.5mm]
$M_1 = \lambda^{13} \Lambda_G$
& $5.656 \times 10^{7} \, \text{GeV}$
& 1st RH neutrino mass
\\[0.5mm]
$M_2 = \lambda^{12} \Lambda_G$
& $2.571 \times 10^{8} \, \text{GeV}$
& 2nd RH neutrino mass
\\[0.5mm]
$M_3 = \lambda^{11} \Lambda_G$
& $1.169 \times 10^{9} \, \text{GeV}$
& 3rd RH neutrino mass
\\[0.5mm]
$M_4 = \lambda^{10} \Lambda_G$
& $5.312 \times 10^{9} \, \text{GeV}$
& 4th RH neutrino mass
\\[0.5mm]
$M_5 = \lambda^{7} \Lambda_G$
& $4.989 \times 10^{11} \, \text{GeV}$
& 5th RH neutrino mass
\\[0.5mm]
$M_6 = \lambda^{6} \Lambda_G$
& $2.268 \times 10^{12} \, \text{GeV}$
& 6th RH neutrino mass
\\[0.5mm]
$Y_{11} = \lambda^{6.5}$
& $5.318 \times 10^{-5}$
& 11 component of $Y_{\nu}$
\\[0.5mm]
$Y_{12} = \lambda^{6.0}$
& $1.134 \times 10^{-4}$
& 12 component of $Y_{\nu}$
\\[0.5mm]
$Y_{13} = \lambda^{5.5}$
& $2.417 \times 10^{-4}$
& 13 component of $Y_{\nu}$
\\[0.5mm]
$Y_{21} = \lambda^{6.0}$
& $1.134 \times 10^{-4}$
& 21 component of $Y_{\nu}$
\\[0.5mm]
$Y_{22} = \lambda^{5.5}$
& $2.417 \times 10^{-4}$
& 22 component of $Y_{\nu}$
\\[0.5mm]
$Y_{23} = \lambda^{5.0}$
& $5.154 \times 10^{-4}$
& 23 component of $Y_{\nu}$
\\[0.5mm]
$Y_{31} = \lambda^{5.5}$
& $2.417 \times 10^{-4}$
& 31 component of $Y_{\nu}$
\\[0.5mm]
$Y_{32} = \lambda^{5.0}$
& $5.154 \times 10^{-4}$
& 32 component of $Y_{\nu}$
\\[0.5mm]
$Y_{33} = \lambda^{4.5}$
& $1.099 \times 10^{-3}$
& 33 component of $Y_{\nu}$
\\[0.5mm]
$Y_{41} = \lambda^{5.0}$
& $5.154 \times 10^{-4}$
& 41 component of $Y_{\nu}$
\\[0.5mm]
$Y_{42} = \lambda^{4.5}$
& $1.099 \times 10^{-3}$
& 42 component of $Y_{\nu}$
\\[0.5mm]
$Y_{43} = \lambda^{4.0}$
& $2.343 \times 10^{-3}$
& 43 component of $Y_{\nu}$
\\[0.5mm]
$Y_{51} = \lambda^{3.5}$
& $4.994 \times 10^{-3}$
& 51 component of $Y_{\nu}$
\\[0.5mm]
$Y_{52} = \lambda^{3.0}$
& $1.065 \times 10^{-2}$
& 52 component of $Y_{\nu}$
\\[0.5mm]
$Y_{53} = \lambda^{2.5}$
& $2.270 \times 10^{-2}$
& 53 component of $Y_{\nu}$
\\[0.5mm]
$Y_{61} = \lambda^{3.0}$
& $1.065 \times 10^{-2}$
& 61 component of $Y_{\nu}$
\\[0.5mm]
$Y_{62} = \lambda^{2.5}$
& $2.270 \times 10^{-2}$
& 62 component of $Y_{\nu}$
\\[0.5mm]
$Y_{63} = \lambda^{2.0}$
& $4.840 \times 10^{-2}$
& 63 component of $Y_{\nu}$
\\ \hline
\label{Tab:para_E6}
\end{tabular} 
\end{center}
\end{table}

The $E_6\times U(1)_A$ GUT in Table \ref{tb:Field contents} predicts
the six RH neutrino masses $M_\alpha$ $(\alpha=1,2,\cdots,6)$ and the Dirac neutrino Yukawa couplings $Y_{\alpha i}$ $(i=1,2,3)$ as in
Table \ref{Tab:para_E6} except $O(1)$ coefficients. 
In the followings, we briefly review the derivation
of these predictions from the model. See Ref.\cite{E6, gauge} for the detail.
 The masses of the RH neutrinos can be obtained through the higher dimensional interactions
\begin{equation}
(Y^{\bar X\bar Y})_{ij}\Psi_i\Psi_j\bar X\bar Y, \quad (\bar X,\bar Y=\bar H,\bar C),
\end{equation}
after developing the VEVs 
$\langle \bar H\rangle\sim \lambda^{-\frac{1}{2}(h+\bar h)}$ and 
$\langle \bar C\rangle\sim \lambda^{-\frac{1}{2}(c+\bar c)}$. (These VEVs are determined
by the VEV relations for the GUT singlet operators $\bar HH$ and $\bar CC$ and 
the $D$-flatness conditions.)
For example, the mass of $S_1$ ($N_{R1}^c$) becomes 
$\lambda^{2\psi_1+2\bar h-(h+\bar h)}\Lambda\sim\lambda^{13}\Lambda$
($\lambda^{2\psi_1+2\bar c-(c+\bar c)}\Lambda\sim\lambda^{12}\Lambda$).
It is convenient to define the effective $U(1)_A$ charges for any fields $\Psi$ as
\begin{equation}
\tilde \psi=\psi+\frac{1}{5}c_V(\Psi)+\frac{1}{2}c_{V'}(\Psi),
\label{effectivecharge}
\end{equation}
where $c_V(\Psi)$ and $c_{V'}(\Psi)$ are the $U(1)_V$ and $U(1)_{V'}$ charges of $\Psi$,
respectively. The coefficients in the above equation (\ref{effectivecharge})
are determined so that
the relations $\langle H\rangle\sim\lambda^{-\tilde h}$,
$\langle \bar H\rangle\sim\lambda^{-\tilde {\bar h}}$,
$\langle C\rangle\sim\lambda^{-\tilde c}$, and
$\langle \bar C\rangle\sim\lambda^{-\tilde {\bar c}}$ are satisfied.
It is obvious that the relations (\ref{SUSYzero}) and (\ref{VEV}) do not change when the effective 
$U(1)_A$ charges are introduced because the $U(1)_V$ 
and $U(1)_{V'}$ charges of the $E_6$ invariant terms are vanishing.
Although special relations between $O(1)$ coefficients due to the $E_6$ symmetry (or other
original symmetries which are broken in the effective theory) cannot be seen explicitly
in the effective model, the effective $U(1)_A$ charges are useful to estimate the
couplings of any terms allowed by the original symmetry. For example, the Dirac Yukawa 
couplings are easily estimated by these effective $U(1)_A$ charges as 
$(\lambda^{\tilde l_i+\tilde n_{Rj}^c+\tilde h_u},\lambda^{\tilde l_i+\tilde s_j+\tilde h_u})$. 
The mass matrices of the RH neutrinos are also calculated as 
$\left(\begin{array}{cc}
 \lambda^{\tilde n_{Ri}^c+\tilde n_{Rj}^c} &  \lambda^{\tilde n_{Ri}^c+\tilde s_j} \\
  \lambda^{\tilde s_i+\tilde n_{Rj}^c}     &   \lambda^{\tilde s_i+\tilde s_j}
\end{array}\right)$. 
The RH neutrino masses $M_\alpha$ in Table \ref{Tab:para_E6} can be obtained by diagonalizing 
the $6\times 6$ RH neutrino mass matrix. 
In Table \ref{Tab:para_E6} we change 
the ordering of the RH neutrinos' generation number $\alpha$ so that
smaller number RH neutrino has smaller mass. The Dirac neutrino Yukawa
couplings in Table \ref{Tab:para_E6} use this new index $\alpha$. 
Even effective higher dimensional interactions which give the light neutrino masses can
be estimated as
\begin{equation}
\lambda^{\tilde l_i+\tilde l_j+2\tilde h_u}l_il_jH_u^2,
\end{equation}
which are also derived from the RH neutrino mass matrix and the Dirac neutrino
Yukawa matrix by the seesaw mechanism. 

One of the most interesting features in the anomalous $U(1)_A$ models is that
the higher dimensional interactions give the same contributions to interactions as
the lower dimensional interactions. For example, the coefficients of Yukawa interactions
$\Psi_i\Psi_jH$ are determined by their total $U(1)_A$ charge as 
$\lambda^{\psi_i+\psi_j+h}$ except $O(1)$ coefficient. The higher dimensional interactions
$\Psi_iA\Psi_jH$, whose coefficients are also determined by the total charge as
$\lambda^{\psi_i+\psi_j+h+a}$, also contribute to the Yukawa interactions $\Psi_i\Psi_jH$
after developing the VEV $\langle A\rangle\sim\lambda^{-a}$ which breaks $SU(5)$ into 
the SM gauge group. The coefficients from the higher dimensional interactions are 
estimated as $\lambda^{\psi_i+\psi_j+h+a}\langle A\rangle\sim \lambda^{\psi_i+\psi_j+h}$,
which is nothing but the coefficients of the original Yukawa interactions except $O(1)$
coefficients. Therefore, the unrealistic GUT relations of Yukawa couplings, for example,
$Y_d=Y_e^t$, can be naturally avoided in the anomalous $U(1)_A$ GUT models because the
higher dimensional interactions with the adjoint Higgs $A$ have different contributions
to the down-type Yukawa couplings from the charged lepton Yukawa couplings after 
developing the VEV of $A$.

\section{Possible enhancement for the right-handed neutrino masses}\label{Sec:enhancement}
It is plausible to enhance a coefficient of an interaction if there are a lot of higher
dimensional interactions which contribute to the coefficient by the same order after
developing the VEVs of the negatively charged operators. Roughly, if there are $N$
higher dimensional interactions which give the same contribution to an interaction,
the enhancement factor can be expected to be $\sqrt{N}$ according to the 
random walk theory. 
Since we have introduced several negatively charged singlets as well as the GUT Higgs 
fields, the number $N$ can be large if the total $U(1)_A$ charge of an interaction
is large. For example, in a simplified model in which all negatively charged fields $\Theta_i$ $(i=1,2,\cdots, n)$ have the $U(1)_A$ charges $\theta_i=-1$, the number of the independent interactions with total $U(1)_A$ charge $c$ is given by $N_n(c)=\frac{(n+c-1)!}{c!(n-1)!}$. 
This number $N_n(c)$ becomes easily large when $c$ and $n$ are large. 
For example,  we obtain that $N_{5}(5)=126$, $N_{5}(10)=1001$, $N_{10}(10)=92378$, $\cdots$.  
In this section, we will show it is plausible that the 1st, 2nd and 3rd 
smallest RH neutrino masses
are enhanced and this enhancement does not change the physical predictions for the light
neutrino sector so much.  

The interactions which contribute to the masses of the RH neutrinos $S_i$ and $N_{Ri}^c$ ($i=1,2,3$) are
$\Psi_i\Psi_i\bar H\bar H$ and $\Psi_i\Psi_i\bar C\bar C$, respectively.
The total $U(1)_A$ charges of these interactions are
(11, 9, 5) for $S_i$ and (9, 7, 3) for $N_{Ri}^c$, while the masses expected by
the symmetry are $(\lambda^{13}, \lambda^{11}, \lambda^{7})$ and 
$(\lambda^{12}, \lambda^{10}, \lambda^{6})$, respectively. This means that the enhancement 
factors $\eta_{S_i}$ and $\eta_{N_{Ri}^c}$ for their masses are expected to be the largest for
the lightest RH neutrino $S_1$, the second largest for the second and the third lightest neutrinos $N_{R1}^c$ and $S_2$.  

In this paper, we do not count the total number of the independent interactions which give the mass term of these RH neutrinos
in the explicit $E_6$ GUT model in Table \ref{tb:Field contents}.
However, we discuss what happens when some of the RH neutrinos have larger masses than those expected
by the symmetry.  It is an important observation that each RH neutrino gives the same order of the contribution 
to all components of the light neutrinos' mass matrix 
$M_{\nu}=Y_{\nu_D}^tM_{\nu_R}^{-1}Y_{\nu_D}\langle H_u\rangle^2$ if its mass is nothing but the value expected by the symmetry. 
Therefore, if one of the enhancement factors $\eta_{S_i}$ and $\eta_{N_{Ri}^c}$ is around one, all components of $M_\nu$ becomes the values expected by the symmetry, and so are all components of the
 diagonalizing matrix.
 In order to obtain three eigenvalues expected by the symmetry, three
 of the six enhancement factors must be around one.
Then all predictions on the light neutrino sector become the same order
as the predictions
without the enhancement factors. Since the lightest neutrino mass has  been fixed only its upper limit by experiments, 
the prediction for it can be different from the predicted value without any enhancement factor. 
Therefore, it is sufficient that two RH neutrinos have their masses which are determined by the symmetry 
for consistency with the present constraints obtained by neutrino experiments.

It looks not to be fair that we consider these enhancement effects only for the RH neutrino masses, 
although the mass terms have much larger $U(1)_A$ charges than the other terms like Yukawa terms. We should
change the $U(1)_A$ charge assignment in Table \ref{tb:Field contents}, when such enhancement effects are 
taken into account.  This subject is beyond the scope of this paper. Here we should emphasize that even after
changing these $U(1)_A$ charges, the mass terms of $S_1$, $S_2$, and 
$N_{R1}^c$ have still much larger $U(1)_A$ charges, and therefore, some
enhancements for their masses are expected. 

The next important question is how large enhancement factor is needed for
sufficient leptogenesis in this $E_6$ GUT model. In the next section, 
we try to answer this question.

%%%%%%%%%%%%%%%%%%%%%%%%%%%%%%%%%%%%%%%%%%% 
\section{Leptogenesis in the $E_6 \times U(1)_A$ model}  \label{Sec:leptoE6}  %%%%%
%%%%%%%%%%%%%%%%%%%%%%%%%%%%%%%%%%%%%%%%%%

In the thermal leptogenesis scenario, thermally produced RH
 neutrinos go 
out of equilibrium as temperature decreases to their mass scale, and their 
CP asymmetric decays produce lepton 
asymmetry~\cite{Fukugita:1986hr}. The lepton asymmetry is converted 
to the baryon asymmetry via the nonperturbative $B+L$ violating sphaleron 
processes~\cite{Kuzmin:1985mm}.

\begin{figure}[t!]
\begin{center}
\includegraphics[width=120mm]{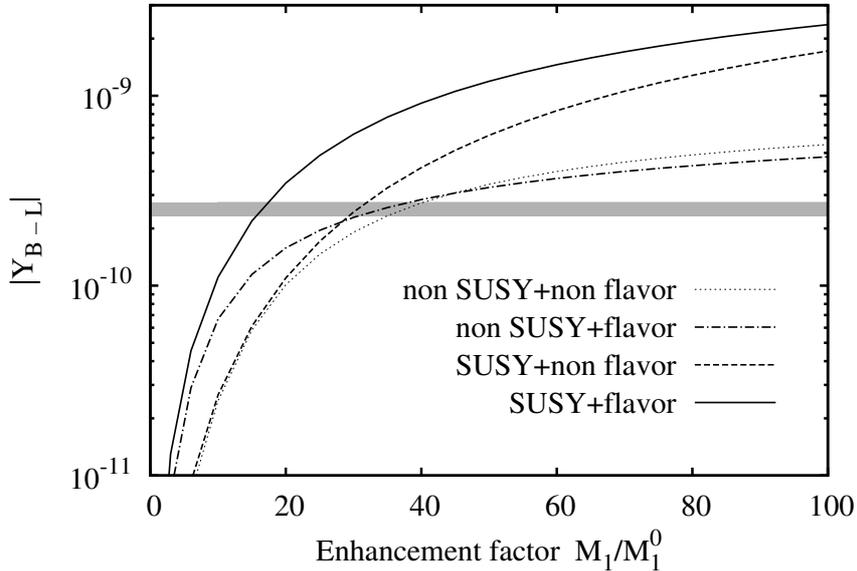}
\caption{$\eta_1\equiv M_{1}/M_1^0$ dependence of $|Y_{B-L}|$ in 
each case. 
Horizontal band corresponds to the observed baryon asymmetry in 
SUSY cases. $M_{1}^{0}$ is the ``bare" Majorana mass in the 
absence of the $U(1)_{A}$ interactions.  
%%%
%%%
We take the simplified CP asymmetry 
$\epsilon_{1}^\text{SM}$\eqref{Eq:simple_eps_SM}, 
$\epsilon_{1i}^\text{SM}$~\eqref{Eq:simple_eps_SM+flavor},  
$\epsilon_{1}^\text{SUSY} = 2 \times \epsilon_{1}^\text{SM}$, and 
$\epsilon_{1i}^\text{SUSY} = 2 \times \epsilon_{1i}^\text{SM}$ 
with the assumption $\Im[(Y^{\dagger} Y)_{61}] = 
\Re[(Y^{\dagger} Y)_{61}]$ for the calculation in each case, 
respectively. $K^{\rm SM}$ and $K^{\rm SUSY}$ can be written as 
$K^{\rm SM}\sim 37/\eta_1$ and $K^{\rm SUSY}\sim 51/\eta_1$, respectively. }
\label{Fig:M1dependence_allcase}
\end{center}
\end{figure}

In this section, we calculate the thermally produced lepton number in the 
$E_6\times U(1)_A$ model with the Dirac neutrino Yukawa couplings $Y_{\alpha i}$ 
$(\alpha=1,2,\cdots, 6, i=1,2,3)$ which are determined 
by the symmetry as in Table \ref{Tab:para_E6} and the masses $M_\alpha$ 
for the mass eigenstate of the RH neutrinos $N_\alpha$. 
Some of six $M_\alpha$ have enhancement factors
$\eta_\alpha$ larger than 1. 
What we would like to know by this calculation is how large enhancement factors are required to
obtain sufficiently large lepton number.  
In the calculation, it is important to
include supersymmetric contributions and the effects of lepton flavor in the
final state of the decay process simultaneously.
To show this statement, we calculate the sufficient enhancement factor in
four cases:
\begin{itemize}
 \vspace{-1.5mm}
 \item non-SUSY + non flavor  
 \vspace{-1mm}
 \item non-SUSY + flavor
 \vspace{-1mm} 
 \item SUSY + non flavor
 \vspace{-1mm}
 \item SUSY + flavor
 \vspace{-1.5mm}
\end{itemize}
The result is shown in Fig. \ref{Fig:M1dependence_allcase}. In a realistic situation of the 
$E_{6} \times U(1)_{A}$ GUT model, i.e., in the case of SUSY+flavor, 
the sufficient lepton number can be obtained if the enhancement factor for the $N_1$ mass is 
around 16. This means that $M_{1}\sim 9\times 10^8$ GeV. 

It is known that supersymmetric contribution is important when $K$ is smaller than 
1 because supersymmetric calculation makes $K$ larger effectively.
On the other hand, the lepton flavor effects are important when the decay parameter 
$K \equiv \Gamma_{N_{1}}(T=0)/H(T=M_{1})$ is larger than 1 because $K$ for the 
muon and the electron become smaller
than $K$ for the tau.  Here $T$ is temperature of the universe. 
Our calculation shows that it is important to inculude both contributions when $K$ is larger 
than 1. This is because supersymmetric contribution is important for smaller $K$ of the 
electron and the muon.

%%%%%%%%%%%%%%%%%%%%%%%%%%%%%%%%%%%%%%%%%%% 
\subsection{non-SUSY + non-flavor}  \label{Sec:nonSUSY+nonflavor}  %%%%%%%%%%
%%%%%%%%%%%%%%%%%%%%%%%%%%%%%%%%%%%%%%%%%%%

\begin{figure}[t!]
\begin{center}
\includegraphics[width=137mm]{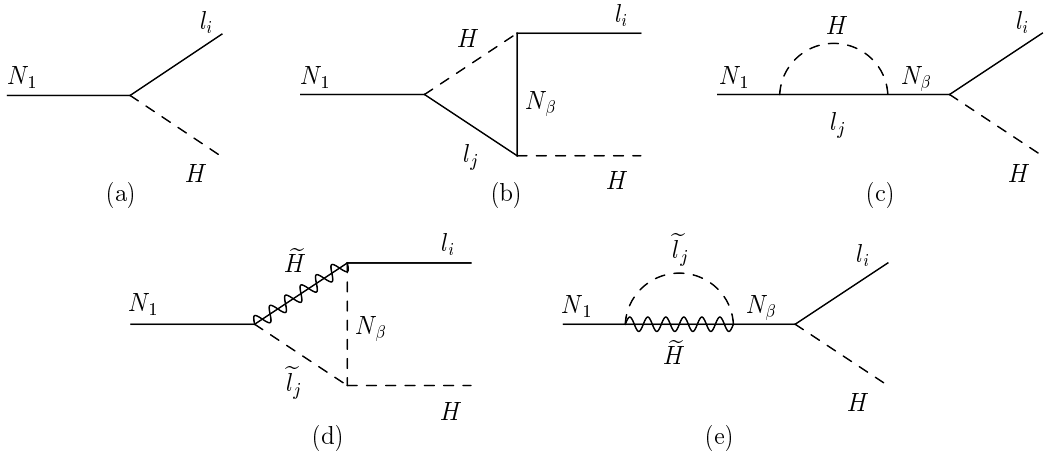}
\caption{Tree and one-loop diagrams contributing to the CP 
asymmetric decay of lightst RH neutrino. In the model, five 
RH neutrinos $N_{\beta} \, (\beta = 2, ..., 6)$ contribute 
to the CP asymmetry. }
\label{Fig:diagram}
\end{center}
\end{figure}

\begin{figure}[t!]
%\begin{minipage}{0.5\hsize}
\begin{center}
\includegraphics[width=100mm]{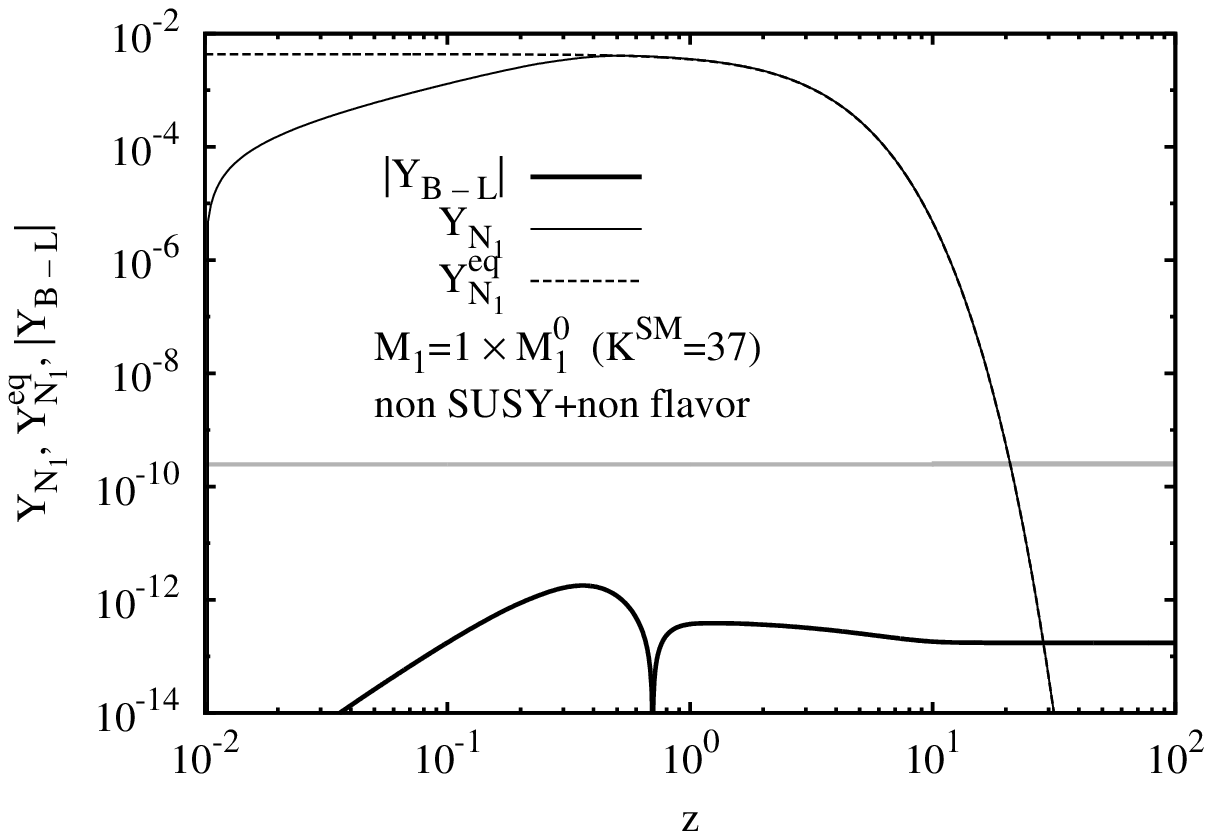}
\end{center}
%\end{minipage}
%\begin{minipage}{0.5\hsize}
\begin{center}
\includegraphics[width=100mm]{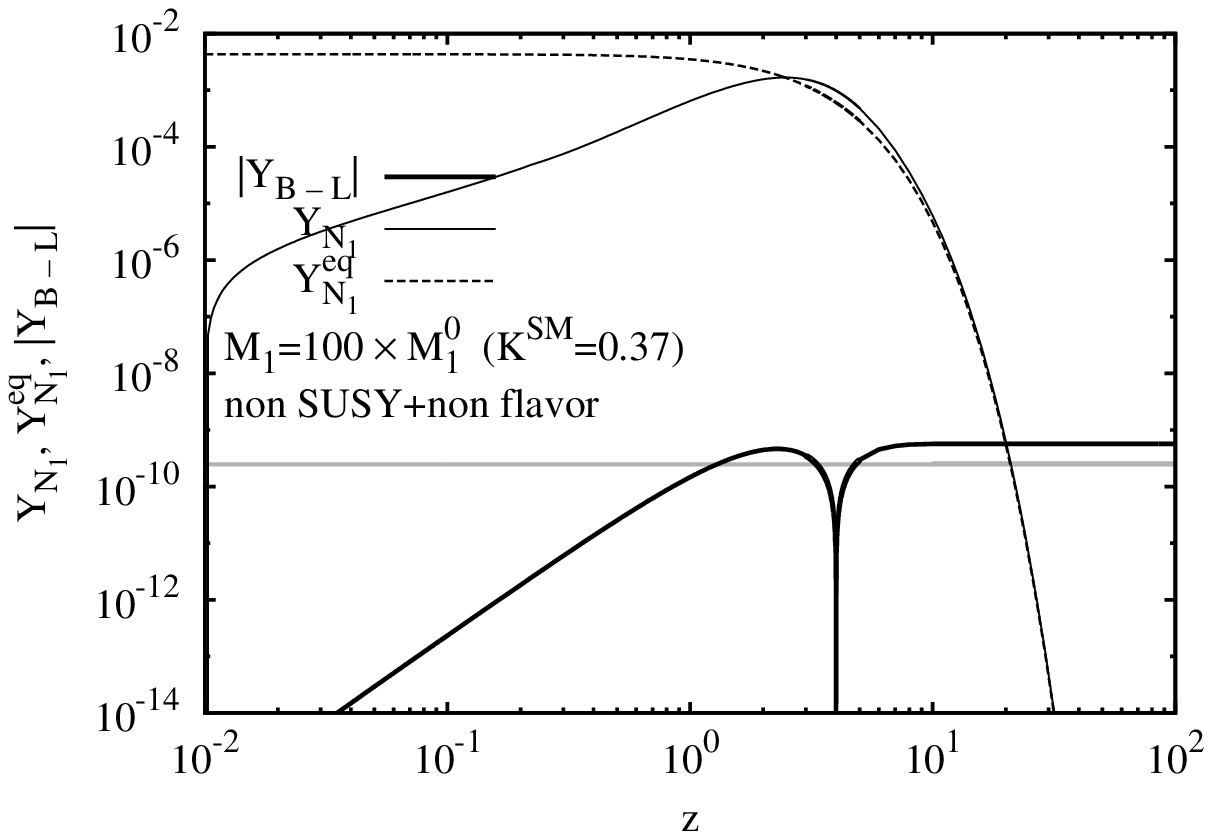}
\end{center}
%\end{minipage}
\caption{Evolutions of $|Y_{B-L}|$, $Y_{N_{1}}$, and 
$Y_{N_{1}}^{eq}$ for $M_{1} = 1 \times M_{1}^{0}$ (top 
panel) and $M_{1} = 100 \times M_{1}^{0}$ (bottom panel) in the 
non-SUSY+non-flavor case. Here $M_{1}^{0}$ is the ``bare" 
Majorana mass in the absence of the $U(1)_{A}$ interactions. 
%%%
%%%
Horizontal band corresponds to the observed baryon asymmetry. We 
take the simplified CP asymmetry \eqref{Eq:simple_eps_SM} 
with the assumption $\Im[(Y^{\dagger} Y)_{61}] = 
\Re[(Y^{\dagger} Y)_{61}]$. }
\label{Fig:evolution_eachM1_nonSUSY_nonflavor}
\end{figure}

In this subsection, we evaluate the lepton asymmetry in the non-SUSY+non-flavor case in 
$E_6\times U(1)_A$ models after brief review. 
This simple calculation is 
important to understand the outline of the thermal leptogenesis with
the RH neutrino masses and the Dirac Yukawa couplings in
Table~\ref{Tab:para_E6}.

In the model, since RH neutrinos are hierarchical in mass, the 
lepton asymmetry is generated by the CP asymmetric reactions of the 
lightest RH neutrino $N_1$.  In the followings, we assume that the lightest
 RH neutrino is $N_1$ while it has the largest enhancement factor.
%%%
%%%
The lepton asymmetry is evaluated by a coupled set of evolution 
equations of the lightest RH neutrino $N_{1}$ and the lepton 
asymmetry $L$: 
%%%
%%%
\begin{equation}
\begin{split}
   \frac{dY_{N_1}}{dz} 
   &= 
   -\frac{z}{sH(z=1)} 
   \left( \frac{Y_{N_1}}{Y_{N_1}^{eq}} - 1 \right)
   \left[ \gamma_{D} + 2\gamma_{Ss} 
   + 4\gamma_{St} \right], 
\label{Eq:N_Boltz}   
\end{split}      
\end{equation}
%%%
%%%
\begin{equation}
\begin{split}
   \frac{dY_L}{dz}
   = 
   -\frac{z}{sH(z=1)} 
   \left\{
   \left[ \frac{1}{2} \frac{Y_{L}}{Y^{eq}_{l}} + \epsilon_{N_{1}}^\text{SM}  
   \left(1-\frac{Y_{N_1}}{Y^{eq}_{N_1}}\right) \right] \gamma_D 
   + \frac{Y_{L}}{Y_{l}^{eq}} \left[2\gamma_{St} + 
   \frac{Y_{N_1}}{Y_{N_1}^{eq}} \gamma_{Ss} \right]
   \right\}. 
   \label{Eq:L_Boltz_nonSUSY_nonflavor}
\end{split}      
\end{equation}
%%%
%%%
Here $s$ is the entropy density, and $H$ is the Hubble parameter. We 
use a dimensionless variable $z \equiv M_{1}/T$. 
$Y_{X}$ and $Y_{X}^{eq}$ are yield value and its equilibrium one of a species $X$, 
respectively, which are the number density normalized to the entropy density. 
%%%
%%%
$\gamma_{D} = \gamma_{D}(N_{1} \leftrightarrow l H)$, 
$\gamma_{Ss} = \gamma_{Ss}(N_{1} l \leftrightarrow Q_{3} T_R^c)$, 
and $\gamma_{St} = \gamma_{St}(N_{1} Q_{3} \leftrightarrow l T_R)$ 
are thermal averaged decay rate (inverse decay rate), s-channel, and 
t-channel scattering rate, respectively~\cite{Plumacher:1997ru}. 
Here $l$, $Q_{3}$, and $T_R$ denote $SU(2)_L$ doublet lepton, third 
generation doublet quark, and singlet top quark, respectively. 
%%%
%%%
The $CP$ asymmetry $\epsilon_{N_{1}}^\text{SM}$ is defined as 
$\epsilon_{N_{1}}^\text{SM} 
= [\Gamma(N_{1} \to l H) - \Gamma(N_{1} \to \bar{l} H^{\dagger})]/
[\Gamma(N_{1} \to l H) + \Gamma(N_{1} \to \bar{l} H^{\dagger})]$. 
The first non-zero contribution to $\epsilon_{N_{1}}^{\rm SM}$ comes 
from interference between tree-level amplitude with the one-loop 
contributions (upper three diagrams in Fig.~\ref{Fig:diagram}), 
and it is 
calculated in a hierarchical limit in RH neutrino masses as 
$\epsilon_{N_{1}}^\text{SM} = -(3/16\pi) 
%\displaystyle{
\sum_{\beta \neq 1}^{6}
%} 
\left( 
\Im \bigl[ \left( Y^{\dagger} Y \right)_{\beta 1}^{2} \bigr]/
\left[ Y^{\dagger} Y \right]_{11} 
\right)
\left( M_{1}/M_{\beta} \right)
$~\cite{Covi:1996wh}.  
Note that, the $E_6$ GUT model has  six RH neutrinos, and therefore, $\beta=2,3, \cdots, 6$.

Key ingredients for the lepton asymmetry generation are the CP asymmetry 
$\epsilon_{N_{1}}^\text{SM}$ and the decay parameter 
$K \equiv \Gamma_{N_{1}}(T=0)/H(T=M_{1})$~\cite{Kolb} which
parametrizes the departure from the thermal equilibrium of RH neutrinos
at $T=M_1$.
$K$ is important because it is related with $\gamma_D$
and the factor $(1-Y_{N_{1}}/Y_{N_{1}}^\text{eq})$ in 
Eq.~\eqref{Eq:L_Boltz_nonSUSY_nonflavor}. 
%%%
%%%
The lepton asymmetry $Y_L$ is essentially determined by the above two 
parameters as
\begin{equation}
Y_L\sim \epsilon^{\rm SM}_{N_1}C(K).
\end{equation}
The behavior of the function $C(K)$ is as follows. 
When $K>1$, $C(K)$ becomes a decreasing function of $K$.
$K>1$ means that the RH neutrinos are still in 
the thermal equilibrium at $T=M_1$, and therefore, the number density of
$N_1$ decreases rapidly when $T<M_1$. This reduces the produced lepton
asymmetry. Obviously larger $K$ results in lower decoupling temperature, smaller $Y_{N1}$ 
after the decoupling and smaller lepton asymmetry.
When $K<1$, $C(K)$ becomes a increasing function of $K$.
$K<1$ means that
the RH neutrinos are out of the thermal equilibrium at $T=M_1$, and 
therefore, the number of thermally produced RH neutrinos becomes smaller
for smaller $K$.  This reduces the produced lepton asymmetry.
Around $K\sim 1$, the function $C(K)$ becomes maximal.
Sufficient lepton asymmetry can be obtained when $K\sim 1$ and
$\epsilon^{\rm SM}_{N_1}\sim 10^{-7}$.

Let us calculate the above two important parameters in the 
$E_6\times U(1)_A$ GUT.
First, we estimate $\epsilon^{\rm SM}_{N_1}$ and $K^{\rm SM}$ without
any enhancement factor $\eta_\alpha$ as
\begin{eqnarray}
	\epsilon_{N_{1}}^\text{SM} 
	&= &
	2 \left(
		- \frac{3}{16 \pi} 
		\frac{\Im \bigl[ \left( Y^{\dagger} Y \right)_{61}^{2} \bigr]}
		{\left[ Y^{\dagger} Y \right]_{11}} 
		\frac{M_{1}}{M_{6}}
	\right) 
	\sim -8.85\times 10^{-9} \left( \frac{M_{1}}{5.7 \times 10^{7} \, \text{GeV}} \right), 
\label{Eq:simple_eps_SM} \\
   K^\text{SM} 
  & = &
   \frac{[Y^{\dagger}Y]_{11} M_{1}/8\pi}
   {1.66 (g_{*}^\text{SM})^{1/2} M_{1}^{2}/M_\text{pl}} 
   \simeq 
   37 \left( \frac{5.7 \times 10^{7} \, \text{GeV}}{M_{1}} \right),
   \label{Eq:K_nonSUSY_nonflavor}
\end{eqnarray}
where $M_\text{pl}$ is the Planck scale and 
$g_{*}^\text{SM}$ is the effective relativistic degrees of freedom, 
which is obtained as $g_{*}^\text{SM} = 106.75$ with the SM particle 
contents.
For the estimation of $\epsilon^{\rm SM}_{N_1}$, we have adopted
two assumptions. The first assumption is that 
$\Im \bigl[ \left( Y^{\dagger} Y \right)_{\beta 1}^{2} \bigr]$ can be
estimated by  
$\Re \bigl[ \left( Y^{\dagger} Y \right)_{\beta 1}^{2} \bigr]$.
This assumption is reasonable because we regard all Yukawa couplings 
as complex numbers.
The second assumption is on the factor 2 in front of the parenthesis in 
Eq. (\ref{Eq:simple_eps_SM}). An important observation is that
$\left(Y^{\dagger} Y \right)_{\beta 1}^{2} 
		\left[ Y^{\dagger} Y \right]_{11}^{-1}
		\frac{M_{1}}{M_{\beta}}\sim \lambda^{11}$ is not dependent on 
		$\beta$. 
Therefore, we can expect an enhancement factor after summation of 
the index $\beta$, and we assume that the enhancement factor is 
two through all calculations in this paper.

It is obvious that the lepton asymmetry with these parameters are too
small to explain the observation.  $K^{\rm SM}$ is too large and
$\epsilon^{\rm SM}_{S_1}$ is too small.  However, as discussed in the 
previous section, the lightest RH neutrino mass can be expected to have
an enhancement factor which can be much larger than one. Interestingly,
when the lightest RH neutrino mass $M_1$ becomes larger, the produced
lepton asymmetry becomes larger because the CP asymmetry $\epsilon^{\rm SM}_{N_1}$ 
becomes larger and  the decay parameter 
$K^{\rm SM}$ becomes smaller as seen in Eq.~\eqref{Eq:simple_eps_SM}.
 For example, if we take the enhancement
factor is around 37, the sufficient lepton asymmetry can be expected
because $K^{\rm SM}\sim 1$ and 
$\epsilon^{\rm SM}_{N_1}\sim 3\times 10^{-7}$.

Figure~\ref{Fig:evolution_eachM1_nonSUSY_nonflavor} shows the 
evolutions of the lepton asymmetry $|Y_{B-L}|$, yield value of the RH 
neutrino $Y_{N_{1}}$, and its equilibrium one $Y_{N_{1}}^{eq}$ 
for $M_{1} = 1 \times M_{1}^{0}$ (top panel) and $M_{1} = 100 
\times M_{1}^{0}$ (bottom panel). Here $M_{1}^{0}$ represents 
the ``bare" Majorana mass, that is the physical mass of the lightest 
RH neutrino without any enhancement factor.
%%%
%%%
The lepton asymmetry for $M_{1}=1 \times M_{1}^{0}$,
$|Y_{L}| \simeq 10^{-13}$,  is too small 
to account for the observed baryon asymmetry. In non-SUSY cases, the 
required lepton asymmetry is $2.285 \times 10^{-10} \leq Y_{B-L} \leq 
2.685 \times 10^{-10}$ with the conversion rate of the lepton asymmetry 
to the baryon asymmetry $Y_{B}=(28/79) Y_{B-L}$~\cite{Harvey:1990qw} 
and the observed baryon number $8.097 \times 10^{-11} \leq Y_{B} 
\leq 9.518 \times 10^{-11}$~\cite{Agashe:2014kda}.
%%%
%%%
For $M_{1}=100 \times M_{1}^{0}$, the lepton asymmetry is drastically 
enhanced. 
The enhancement of the lightest RH neutrino mass makes the CP asymmetry larger 
[see Eq.~\eqref{Eq:simple_eps_SM}] and reduces the $K^\text{SM}$ factor. 
Indeed, in the bottom panel in Fig.~\ref{Fig:evolution_eachM1_nonSUSY_nonflavor}, 
we find the larger deviation from thermal equilibrium compared with the top 
panel. The combination of these effects leads the enhancement of lepton 
asymmetry.

Dotted line in Fig.~\ref{Fig:M1dependence_allcase} shows the $M_{1}$ 
dependence of the lepton asymmetry in the non-SUSY+non-flavor case. 
Since numerically $C(K)\propto K$ for $K\ll 1$, the lepton asymmetry
becomes asymptotically a constant for the enhancement factor larger
than 40. 
%%%
%%%
In this case, the physical mass of the RH neutrino is required to be $M_{1} 
= (35-39) \times M_{1}^{0}$ to account for the observed baryon number.

%%%%%%%%%%%%%%%%%%%%%%%%%%%%%%%%%%%%%%%%%%% 
\subsection{non-SUSY + flavor}  \label{Sec:NonSUSY_flavor}  %%%%%%%%%%
%%%%%%%%%%%%%%%%%%%%%%%%%%%%%%%%%%%%%%%%%%%

\begin{figure}[t!]
%\begin{minipage}{0.5\hsize}
\begin{center}
\includegraphics[width=100mm]{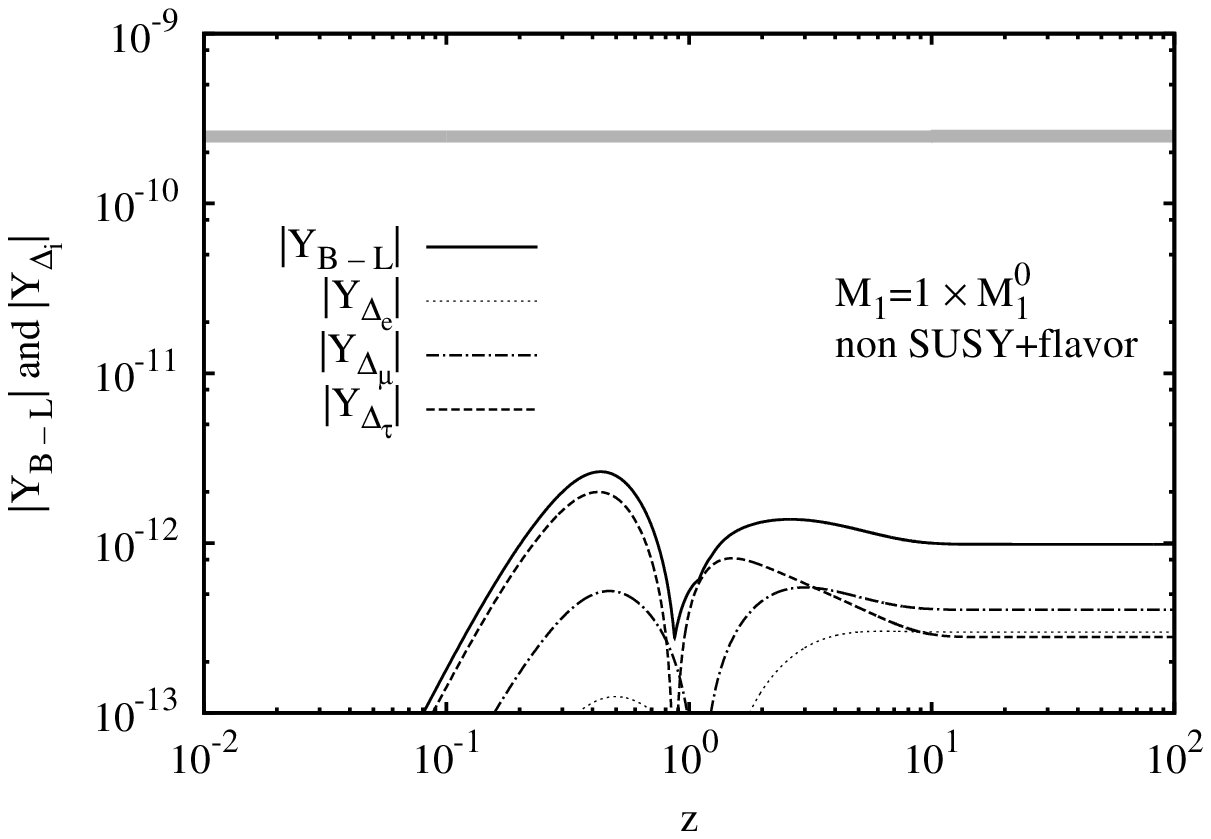}
\end{center}
%\end{minipage}
%\begin{minipage}{0.5\hsize}
\begin{center}
\includegraphics[width=100mm]{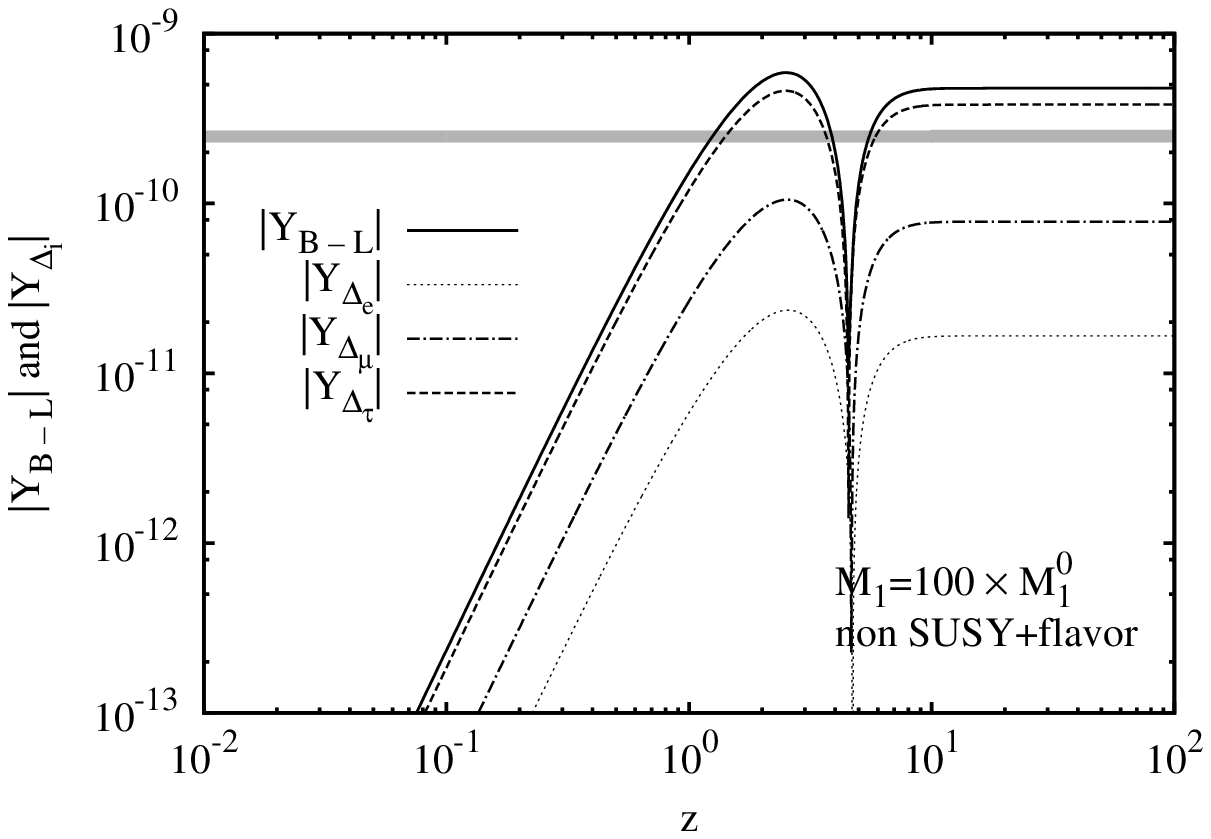}
\end{center}
%\end{minipage}
\caption{Same as Fig.~\ref{Fig:evolution_eachM1_nonSUSY_nonflavor}, 
but in the non-SUSY+flavor case. The evolutions of electron number $L_{e}$, 
muon number $L_{\mu}$, and tau number $L_{\tau}$ are also plotted. 
Since the evolution of RH neutrino is same as the non-SUSY+non-flavor case, 
we cut the part.}
\label{Fig:evolution_eachM1_nonSUSY_flavor}
\end{figure}

It is important for the evaluation of lepton asymmetry in the $E_{6} \times 
U(1)_{A}$ model to separately involve each lepton flavor channel of the CP 
asymmetric decays. The reasons are as follows. 
%%%
%%%
The $E_6$ model possesses the features: 
(i) the evolution of lightest RH neutrino is in the strong washout regime, \ 
(i\hspace{-1pt}i) all of asymmetry productions of each lepton flavor by the 
CP asymmetric decay are sizable. 
%%%
%%%
These features can give rise to $\mathcal{O}(1)$ corrections to the final 
lepton asymmetry with respect to the case where the flavor effects are 
ignored~\cite{Nardi:2006fx}. This is because that the evolutions of the 
lepton asymmetries of each lepton flavor are in the regime of washout with 
different magnitudes. In this section we briefly review the leptogenesis with 
the flavor effects.

The total lepton asymmetry is given by the sum of the asymmetry of each 
lepton flavor, $Y_{B-L} = Y_{\Delta_{e}} + Y_{\Delta_{\mu}} + 
Y_{\Delta_{\tau}}$, where $\Delta_{i} = B/3 - L_{i}$. The subscript $i$ 
represents the lepton flavor. The evolutions of each asymmetry is described 
by the flavor dependent Boltzmann equations, 
%%%
%%%
\begin{equation}
\begin{split}
   &
   \frac{dY_{\Delta_{i}}}{dz} 
   = 
   -\frac{z}{sH(z=1)} 
   \Biggl\{  
   \left( \frac{Y_{N_1}}{Y_{N_1}^{eq}} - 1 \right) 
   \epsilon_{1i}^\text{SM} \gamma_{D} 
   + K_{i}^{0} \sum_{j} 
   \biggl[ 
   \frac{1}{2} \left( C_{ij}^{l} + C_{j}^{H} \right) \gamma_{D} 
   \\& ~~ + 
   \left( \frac{Y_{N_1}}{Y_{N_1}^{eq}} - 1 \right) 
   \left( C_{ij}^{l} \gamma_{S_{s}}  
   + \frac{C_{j}^{H}}{2} \gamma_{S_{t}} \right) 
   + \left( 2 C_{ij}^{l} + C_{j}^{H} \right) 
   \left( \gamma_{S_{t}} + \frac{\gamma_{S_{s}}}{2} \right)
   \biggr] 
   \frac{Y_{\Delta_{i}}}{Y_{l}^{eq}}
   \Biggr\}. 
   \label{Eq:Boltz_nonSUSY_flavor}
\end{split}      
\end{equation}
%%%
%%%
The coefficient $K_{i}^{0}$ is the flavor projection, 
$K_{i}^{0} = Y_{1i} Y_{1i}^{*}/(YY^{\dagger})_{11}$. 
%%%
%%%
The flavor dependent CP asymmetry is defined as 
$\epsilon_{1i}^\text{SM} = [\Gamma (N_{1} \to l_{i} H) 
- \Gamma (N_{1} \to \bar{l}_{i} H^{\dagger})]/[\Gamma (N_{1} 
\to l_{i} H) + \Gamma (N_{1} \to \bar{l}_{i} H^{\dagger})]$, 
%%%
%%%
and is calculated in the hierarchical mass limit as 
$\epsilon_{1i}^\text{SM} = 
-\left( 1/8\pi (YY^{\dagger})_{11} \right) 
%\displaystyle{
\sum_{\beta \neq 1}^{6}
%} 
\Im \Bigl\{ 
Y_{\beta i} Y_{1 i}^{*} 
\Bigl[
\left( 3/2 \right) 
\left( M_{1}/M_{\beta} \right) 
\bigl( YY^{\dagger} \bigr)_{\beta 1} 
\\[-1.5mm]
+ 
\left( 
M_{1}^{2}/M_{\beta}^{2}
\right)
\left( 
YY^{\dagger}
\right)_{1 \beta}
\Bigr]
\Bigr\}$~\cite{Covi:1996wh}.  
%%%
%%%
We follow the considerations for deriving Eq.~\eqref{Eq:simple_eps_SM}, 
and obtain the simplified CP asymmetry in the non-SUSY+flavor case as 
%%%
%%%
\begin{equation}
\begin{split}
   \epsilon_{1i}^\text{SM} 
   = 
   2 \left(
   \frac{-1}{8\pi (YY^{\dagger})_{11}}  
   \Im \left\{ 
   Y_{6i} Y_{1 i}^{*} 
   \left[ 
   \frac{3}{2} \frac{M_{1}}{M_{6}} 
   (YY^{\dagger})_{61} 
   + 
   \frac{M_{1}^{2}}{M_{6}^{2}} 
   (YY^{\dagger})_{16}
   \right]
   \right\} 
   \right) . 
   \label{Eq:simple_eps_SM+flavor}
\end{split}      
\end{equation}

The coefficient $C^{l}$ ($C^{H}$) in Eq.~\eqref{Eq:Boltz_nonSUSY_flavor} 
is introduced as the conversion factor between the asymmetry 
normalized to the equilibrium number density for $l_{i}$ ($H$) and 
the yield value of each lepton number normalized to equilibrium 
lepton density as 
$\left( n_{l_{i}} - n_{\bar{l}_{i}} \right)/n_{l_{i}}^\text{eq} = 
- \sum_{j} C_{ij}^{l} \left( Y_{\Delta_{j}}/Y_{l}^\text{eq} \right)$, 
and 
$\left( n_{H} - n_{\bar{H}} \right)/n_{H}^\text{eq} = 
- \sum_{j} C_{j}^{H} \left( Y_{\Delta_{j}}/Y_{l}^\text{eq} \right)$. 
%%%
%%%
The entries are model-independent, which are determined by constraints 
among the chemical potentials enforced by the equilibrium reactions in the 
temperature $T \sim M_{1}$ where the asymmetries are generated. 
%%%
%%%
The region of RH neutrino mass we consider is $1 \leq M_{1}/M_{1}^{0} 
\leq 100$, and the relevant temperature of the leptogenesis in the model 
is in $10^{5} \, \text{GeV} \lesssim T \lesssim 10^{11} \, 
\text{GeV}$. 
%%%
%%%
In this range, SM gauge interactions, third 
and second generation Yukawa interactions are in equilibrium, and the 
equilibrium conditions lead to the following $C^{l}$ and 
$C^{H}$~\cite{Nardi:2006fx}, 
%%%
%%%
\begin{equation}
\begin{split}
   C_{ij}^{l} = \frac{1}{2148}
\begin{pmatrix} 
   906 & -120 & -120
   \\
   -75 & 688 & -28 
   \\
   -75 & -28 & 688 
\end{pmatrix}, \ ~~~ 
    C^{H} = \frac{1}{358} 
\begin{pmatrix} 
   37 & 52 & 52
\end{pmatrix}. 
\end{split}      
\end{equation}

Figure~\ref{Fig:evolution_eachM1_nonSUSY_flavor} shows the evolutions 
of total lepton asymmetry $\left| Y_{B-L} \right|$, and of the asymmetry of 
each lepton flavor $\left| Y_{\Delta_{i}} \right| \ (i = e, \mu, \tau)$ for 
$M_{1} = 1 \times M_{1}^{0}$ (top panel) and for $M_{1} = 100 \times 
M_{1}^{0}$ (bottom panel). 
%%%
%%%
For the interpretation of the result, we need to see both the magnitude of 
the washout and the production efficiency of each lepton asymmetry. 
We rearrange the $K^\text{SM}$ factor~\eqref{Eq:K_nonSUSY_nonflavor} 
to involve the flavor dependence as $K_{i}^\text{SM} = K_{i}^{0} 
K^\text{SM}$. Each $K_{i}^\text{SM}$ is obtained as follows, 
%%%
%%%
\begin{equation}
\begin{split}
	K_{e}^\text{SM} 
	= 
	\frac{\Gamma^\text{SM}_{N_{1} \to l_{e}H} (T=0)}
	{H (T=M_{1})} 
	\simeq 
	1.4 \left( 
	\frac{5.7 \times 10^{7} \, \text{GeV}}
	{M_{1}} 
	\right), 
	\label{Eq:Ke_SM}
\end{split}      
\end{equation}
%%%
%%%
\begin{equation}
\begin{split}
	K_{\mu}^\text{SM} 
	= 
	\frac{\Gamma^\text{SM}_{N_{1} \to l_{\mu}H} (T=0)}
	{H (T=M_{1})} 
	\simeq 
	6.4 \left( 
	\frac{5.7 \times 10^{7} \, \text{GeV}}
	{M_{1}} 
	\right), 
	\label{Eq:Kmu_SM}
\end{split}      
\end{equation}
%%%
%%%
\begin{equation}
\begin{split}
	K_{\tau}^\text{SM} 
	= 
	\frac{\Gamma^\text{SM}_{N_{1} \to l_{\tau}H} (T=0)}
	{H (T=M_{1})} 
	\simeq 
	29 \left( 
	\frac{5.7 \times 10^{7} \, \text{GeV}}
	{M_{1}} 
	\right). 
	\label{Eq:Ktau_SM}
\end{split}      
\end{equation}
%%%
%%%
The $K_{i}^\text{SM}$ is a measure of magnitude of the washout of 
each lepton asymmetry, that is the same with the relation between the 
$K^\text{SM}$ factor and the washout of total lepton asymmetry. 
%%%
%%%
On the other hand, the ratio of asymmetry productions of each 
lepton flavor by the CP asymmetric decay is equal to the ratio of each 
CP asymmetry, and is obtained as follows 
%%%
%%%
\begin{equation}
\begin{split}
	\frac{L_{e} \ \text{production}}{L_{\tau} \ \text{production}} 
	&= 
	\frac{\Im\left[ Y_{61}Y_{11}^{*} \right]}
	{\Im\left[ Y_{63}Y_{13}^{*} \right]}
	= 
	0.048, ~~
	\\
	\frac{L_{\mu} \ \text{production}}{L_{\tau} \ \text{production}} 
	&= 
	\frac{\Im\left[ Y_{62}Y_{12}^{*} \right]}
	{\Im\left[ Y_{63}Y_{13}^{*} \right]}
	= 
	0.220. 
\end{split}      
\end{equation}
%%%
%%%
Here we assumed $\Im\left[ Y_{6i}Y_{1i}^{*} \right] = 
\Re\left[ Y_{6i}Y_{1i}^{*} \right]$. 
For $M_{1} = 1 \times M_{1}^{0}$, $K_{e}$ and $K_{\mu}$ are 
$\mathcal{O}(1)$, but $K_{\tau} \simeq 29$. Each lepton asymmetry 
is generated by the CP asymmetric decays, and then a large part of the 
tau's is washed out by the inverse decays and so on, while the electron's 
and the muon's soon decouple from the equilibrium and survive. 
%%%
%%%
Thus, nonetheless the production efficiencies of the electron and the 
muon number are lower than the tau's, they yields a large part of the 
lepton asymmetry.  
%%%
%%%
On the other hand, for $M_{1} = 100 \times M_{1}^{0}$, $K_{\tau} 
\simeq \mathcal{O}(10^{-1})$, and $K_{e}, \, K_{\mu} \simeq 
\mathcal{O}(10^{-2})$. These $K_{i}$ factors indicate that each 
lepton number generated by the CP asymmetric decay survives without 
being strongly washed out. Thus a large part of lepton asymmetry is 
governed by the tau's.

The chain line in Fig.~\ref{Fig:M1dependence_allcase} shows the $M_{1}$
dependence of the total lepton asymmetry in the non-SUSY+flavor case. 
In the non-SUSY+flavor case, the physical mass of the RH neutrino is required 
to be $30 \leq M_{1}/M_{1}^{0} \leq 37$ to 
account for the observed baryon number. 
%%%
%%%
For $M_{1}/M_{1}^{0} \lesssim 40$, in the non-SUSY+non-flavor case, 
the evolution of the lepton asymmetry is in the strong washout regime. 
While, in the non-SUSY+flavor case, the muon and the 
electron asymmetries are not hardly washed out, and yield sizable contribution 
to total lepton asymmetry. The lepton asymmetry with the flavor effects is 
therefore larger than that in the case where the effects are ignored. 
%%%
%%%
On the other hand, for $M_{1}/M_{1}^{0} \gtrsim 40$, the evolutions 
of the asymmetries of all lepton flavor are in the weak washout regime, 
and hence total lepton asymmetry is determined by only the asymmetry 
production by the CP asymmetric decay. Due to the additional washout 
contributions, in the parameter region, the final lepton asymmetry can be 
smaller than that the case without the flavor effects~\cite{Nardi:2005hs, 
Nardi:2006fx}.

%%%%%%%%%%%%%%%%%%%%%%%%%%%%%%%%%%%%%%%%%%% 
\subsection{SUSY + non-flavor}  \label{Sec:SUSY_nonflavor}  %%%%%%%%%%%%%
%%%%%%%%%%%%%%%%%%%%%%%%%%%%%%%%%%%%%%%%%%%

The SUSY extension of the leptogenesis gives an enhancement for the 
lepton asymmetry which is roughly estimated as
\begin{equation}
\frac{Y_B^{\rm SUSY}}{Y_B^{\rm SM}}\sim \left\{\begin{array}{ll}
                             \sqrt{2} & ( {\rm strong\ washout} ) \\
                             2\sqrt{2} & ({\rm weak\ washout})
                             \end{array}\right.
                             \label{enhancement}
\end{equation}
in Ref. {\cite{Davidson:2008bu}}.
In this section, we briefly review the corrections to interpret the numerical 
results in the context of the $E_{6} \times U(1)_{A}$ model. 

We have two important points which increase the lepton asymmetry. 
The additional decay channels correct 
the definition of the CP asymmetry as 
$\epsilon_{N_{1}}^\text{SUSY} = 
\bigl[ 
\Gamma (N_{1} \to l H) 
- \Gamma (N_{1} \to \bar{l} H^{\dagger}) 
+ \Gamma (N_{1} \to \tilde{l} \tilde{H}) 
- \Gamma (N_{1} \to \tilde{l}^{*} \bar{\tilde{H}})
\bigr]/
\Gamma_{N_{1}}^\text{SUSY}$. 
%%%
%%%
Similarly, for the RH sneutrino, 
$\epsilon_{\tilde N_{1}}^\text{SUSY} = 
\bigl[ 
\Gamma (\tilde{N}_{1} \to \tilde{l} H) 
- \Gamma (\tilde{N}_{1} \to \tilde{l}^{*} H^{\dagger}) 
+ \Gamma (\tilde{N}_{1} \to l \tilde{H}) 
- \Gamma (\tilde{N}_{1} \to \bar{l} \bar{\tilde{H}})
\bigr]/
\Gamma_{\tilde N_{1}}^\text{SUSY}$. 
%%%
%%%
Here $\Gamma_{\tilde N_{1}}^\text{SUSY}$ is the total width 
of the RH sneutrino.  
%%%
%%%
The CP asymmetry receives the contributions of not only the RH neutrinos also 
of its scalar partner, and is obtained as $\epsilon_{N_{1}}^\text{SUSY} = 
-(3/8\pi) 
%\displaystyle{
\sum_{\beta \neq 1}^{6}
%}  
\left( 
\Im \bigl[ \left( Y^{\dagger} Y \right)_{\beta 1}^{2} \bigr]/
\left( Y^{\dagger} Y \right)_{11}
\right)
\left( M_{1}/M_{\beta} \right)$ 
%%%
%%%
in the hierarchical limit of RH neutrino masses~\cite{Covi:1996wh}. The 
CP asymmetry of the RH sneutrino is equal to that of the RH neutrino in 
the hierarchical mass limit. Repeating the consideration for deriving 
Eq.~\eqref{Eq:simple_eps_SM}, we obtain the simplified CP asymmetries 
in the SUSY+non-flavor case as follows, 
%%%
%%%
\begin{equation}
\begin{split}
	\epsilon_{N_{1}}^\text{SUSY} 
	= 
	\epsilon_{\tilde N_{1}}^\text{SUSY} 
	= 
	2 \left( 
		- \frac{3}{8 \pi} 
		\frac{\Im \bigl[ \left( Y^{\dagger} Y \right)_{6 1}^{2} \bigr] } 
		{\left( Y^{\dagger} Y \right)_{11}}
		\frac{M_{1}}{M_{6}}
	\right)
	= 
	2 \times \epsilon_{N_{1}}^\text{SM}. 
	\label{Eq:CP_SUSY}
\end{split}      
\end{equation}
These effects make the lepton asymmetry four times larger.

However, the effective relativistic degrees of freedom 
$g_*^{\rm SUSY}=228.75$ 
is  about twice of $g_*^{\rm SM}$, which reduces the lepton asymmetry to 
entropy ratio by 1/2.

The most important one is the correction of $K$ factor. In the context of 
SUSY, there exists additional decay channels of the RH neutrino, and 
the total width is obtained as 
$\Gamma_{N_{1}}^\text{SUSY} 
= \Gamma (N_{1} \to l H) 
+ \Gamma (N_{1} \to \bar{l} H^{\dagger}) 
+ \Gamma (N_{1} \to \tilde{l} \tilde{H}) 
+ \Gamma (N_{1} \to \tilde{l}^{*} \bar{\tilde{H}}) 
= [Y^{\dagger}Y]_{11} M_{1}/4\pi 
= 2 \times \Gamma_{N_{1}}^\text{SM}$. 
%%%
%%%
$\tilde l$ and $\tilde H$ represent SUSY partners of the $SU(2)_{L}$ 
lepton and the Higgs doublet, respectively. The $K$ factor is calculated 
as follows, 
%%%
%%%
\begin{equation}
\begin{split}
   K^\text{SUSY} 
   = 
   \frac{\Gamma_{N_{1}}^\text{SUSY}(T=0)}{H(T=M_{1})}
   = 
   \frac{[Y^{\dagger}Y]_{11} M_{1}/4\pi}
   {1.66 (g_{*}^\text{SUSY})^{1/2} M_{1}^{2}/M_\text{pl}} 
   \simeq 
   51 \left( \frac{5.7 \times 10^{7} \, \text{GeV}}{M_{1}} \right). 
   \label{Eq:K_SUSY_nonflavor}
\end{split}      
\end{equation}
%%%
%%%
Roughly, the factor $K^\text{SUSY}$ is $\sqrt{2}$ times larger than that in the SM.
This effect reduces the lepton asymmetry in strong washout regime and
enhances it in weak washout regime.

Finally we note the conversion rate from the lepton asymmetry to the 
baryon asymmetry. In the context of SUSY, the additional 
equilibrium reactions at the temperature $T \simeq M_{1}$ alter 
the constraints among the chemical potentials. 
%%%
%%%
The alteration leads the conversion rate as $Y_{B} = 
(8/23) Y_{B-L}$~\cite{Laine:1999wv}. Consequently the required 
lepton asymmetry in SUSY cases is $2.328 \times 10^{-10} \lesssim 
|Y_{B-L}| \lesssim 2.736 \times 10^{-10}$. 

With all these effects, the result in Eq. (\ref{enhancement}) is obtained. 

Dashed line in Fig.~\ref{Fig:M1dependence_allcase} shows the $M_1$ 
dependence of the lepton asymmetry in the SUSY+non-flavor case. The 
lepton asymmetry is given by a sum of partial asymmetries from the 
CP asymmetric decays of RH neutrino and its scalar partner. It is  
evaluated by a coupled set of evolution equations of the RH neutrino, its 
scalar partner, and the partial asymmetries~\cite{Plumacher:1997ru}. 
%%%
%%%
Due to too strong washout, for $M_{1} \lesssim 30 \times M_{1}^{0}$, 
nonetheless the additional CP asymmetric decays, 
the lepton asymmetry is close to that in the non-SUSY+non-flavor case. 
%%%
%%%
While, for $M_{1} \gtrsim 30 \times M_{1}^{0}$, because of both 
$K \simeq \mathcal{O}(1)$ and the additional CP asymmetric decays, 
larger lepton asymmetry is generated than those in the cases of 
non-SUSY+non-flavor and non-SUSY+flavor.

%%%%%%%%%%%%%%%%%%%%%%%%%%%%%%%%%%%%%%%%%%% 
\subsection{SUSY + flavor}  \label{Sec:SUSY_flavor} %%%%%%%%%%%%%%%
%%%%%%%%%%%%%%%%%%%%%%%%%%%%%%%%%%%%%%%%%%%

\begin{figure}[t!]
\begin{center}
\includegraphics[width=120mm]{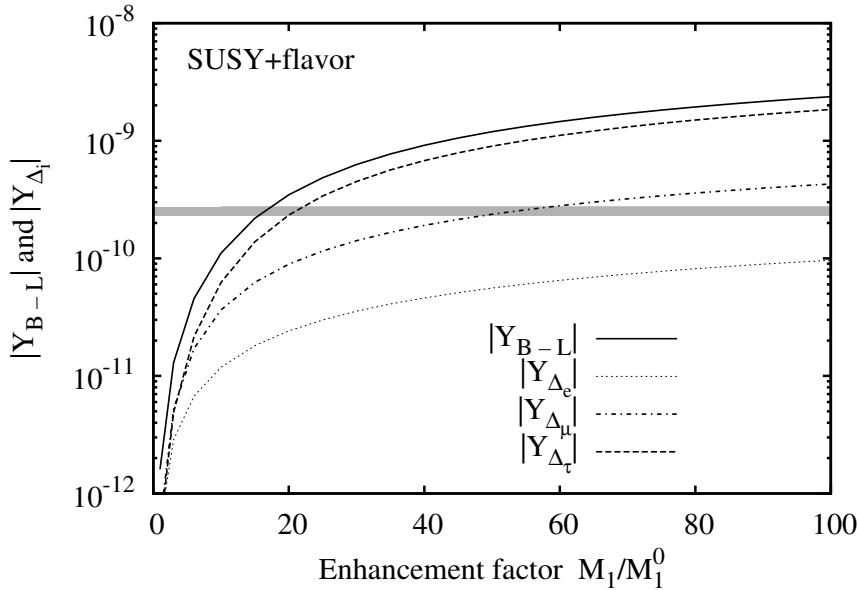}
\caption{$\eta_1\equiv M_{1}/M_1^0$ dependence of $|Y_{B-L}|$ and $|Y_{\Delta_{i}}|$ in 
the SUSY+flavor case. Horizontal band corresponds to the observed baryon 
asymmetry. We take the simplified CP asymmetry $\epsilon_{1i}^\text{SUSY} 
= 2 \times \epsilon_{1i}^\text{SM}$ with the assumption 
$\Im[(Y^{\dagger} Y)_{61}] = \Re[(Y^{\dagger} Y)_{61}]$, where 
$\epsilon_{1i}^\text{SM}$ is given by~\eqref{Eq:simple_eps_SM+flavor}. 
$K^{\rm SUSY}$ can be written as $K^{\rm SUSY}\sim 51/\eta_1$.}
\label{Fig:M1dep_SUSY+flavor}
\end{center}
\end{figure}

\begin{figure}[t!]
\begin{center}
\includegraphics[width=120mm]{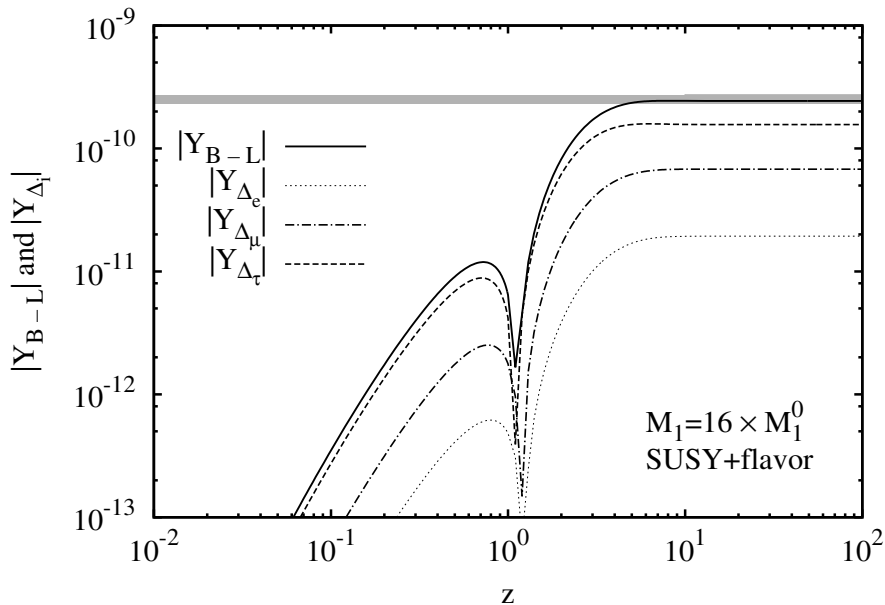}
\caption{Evolutions of $|Y_{B-L}|$ and $|Y_{\Delta_{i}}|$ 
for $M_{1} = 16 \times M_{1}^{0}$ in the SUSY+flavor case. 
Horizontal band corresponds to the observed baryon asymmetry. 
We take the simplified CP asymmetry $\epsilon_{1i}^\text{SUSY} 
= 2 \times \epsilon_{1i}^\text{SM}$ with the assumption 
$\Im[(Y^{\dagger} Y)_{61}] = \Re[(Y^{\dagger} Y)_{61}]$, where 
$\epsilon_{1i}^\text{SM}$ is given by~\eqref{Eq:simple_eps_SM+flavor}. }
\label{Fig:evo_16M0_SUSY+flavor}
\end{center}
\end{figure}

\begin{figure}[t!]
\begin{center}
\includegraphics[width=120mm]{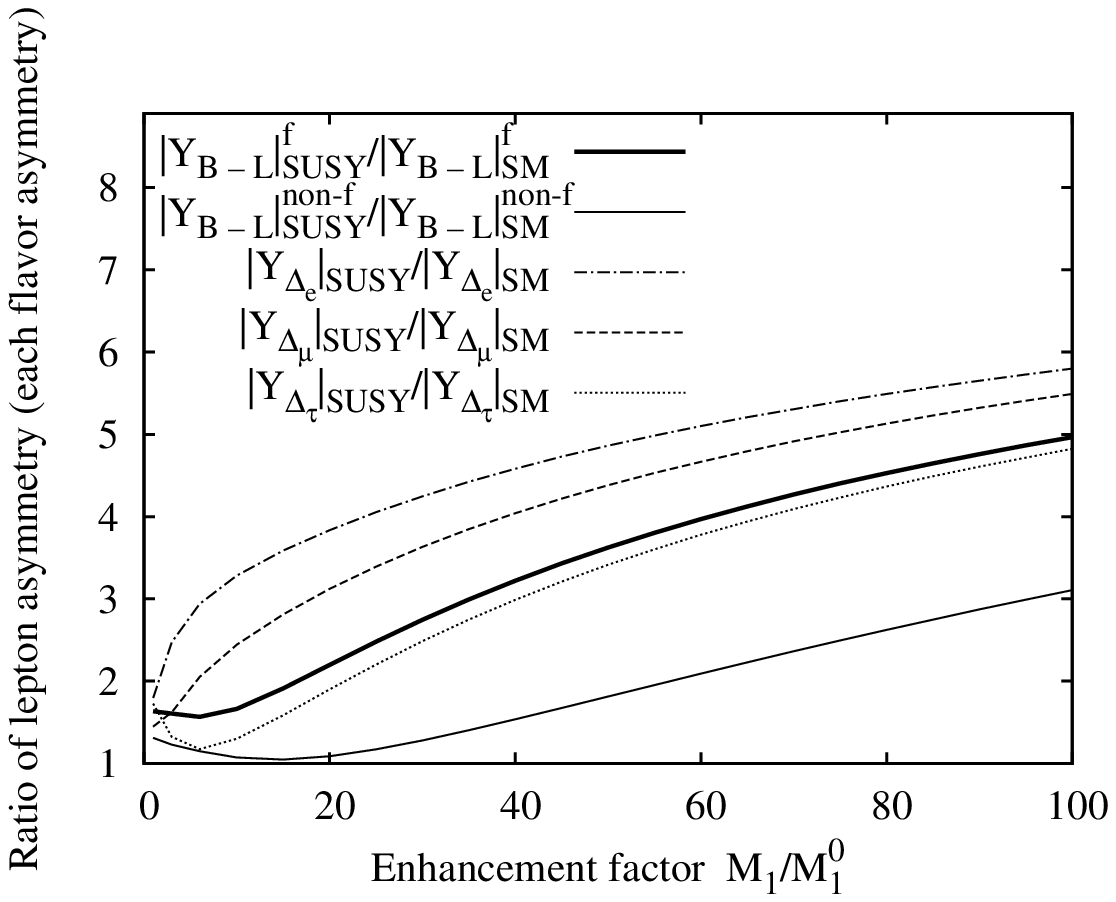}
\caption{$\eta_1\equiv M_{1}/M_1^0$ dependence of the ratio of flavor contributions 
in each case, $\left( |Y_{\Delta_{i}}| \right)_\text{SUSY}/
\left( |Y_{\Delta_{i}}| \right)_\text{SM}$,
$\left( |Y_{B-L}| \right)_\text{SUSY}^{\rm f}/
\left( |Y_{B-L}\right)_\text{SM}^{\rm f}$,
$\left( |Y_{B-L}| \right)_\text{SUSY}^{\rm non-f}/
\left( |Y_{B-L}\right)_\text{SM}^{\rm non-f}$.
Here, for example, $|Y_{B-L}|_{\rm SM}^{\rm f}$ is the lepton asymmetry
in the case non-SUSY+flavor, and $|Y_{B-L}|_{\rm SUSY}^{\rm non-f}$
is the lepton asymmetry in the case SUSY+non-flavor, etc.
$K^{\rm SUSY}$ can be written as $K^{\rm SUSY}\sim 51/\eta_1$.}
\label{Fig:flavor_ratio}
\end{center}
\end{figure}

We are now in a position to discuss the lepton asymmetry in the 
SUSY+flavor case, which involves the SUSY particles contributions 
with the flavor effects [see Sec.~\ref{Sec:NonSUSY_flavor}]. 
This is the realistic situation in the $E_{6} \times U(1)_{A}$ model. 
Interestingly, even in the strong washout regime which is defined by
$K^{\rm SM}>1$, the effect of SUSY
becomes sizable if the flavor effects are included.

The sum of total lepton and slepton asymmetries, $Y_{L}^{(f)}$ and 
$Y_{L}^{(s)}$, converts to the baryon asymmetry, and which are given 
by the sum of the asymmetry of each lepton and slepton flavor: 
$Y_{B-L} = Y_{L}^{(f)} + Y_{L}^{(s)}
= (Y_{\Delta_{e}} 
+ Y_{\Delta_{\mu}} 
+ Y_{\Delta_{\tau}}) 
+ (Y_{\tilde{\Delta}_{e}} 
+ Y_{\tilde{\Delta}_{\mu}} 
+ Y_{\tilde{\Delta}_{\tau}})$. 
%%%
%%%
We take the SUSY spectrum to be $\mathcal{O}(1\,\text{TeV})$ in the 
$E_{6} \times U(1)_{A}$ model. Then, throughout the temperature region 
we consider, the equality of chemical potentials of a SM particle and its 
superpartner, which is refferd as superequilibration~\cite{Chung:2009qs}, 
is maintained in the presence of equilibrium supergauge (or Yukawa) 
reactions.
%%%
%%%
In the superequilibrium regime, since the equality of asymmetries of each 
lepton and its scalar partner is also maintained, 
$Y_{B-L} = 2 \times 
(Y_{\Delta_{e}} 
+ Y_{\Delta_{\mu}} 
+ Y_{\Delta_{\tau}})$. 
%%%
%%%
Thus the baryon asymmetry in the case is evaluated by a coupled set of 
evolution equations of the RH neutrino, its scalar partner, and asymmetry 
of each lepton flavor. The flavor dependent Boltzmann equations are shown in 
Ref.~\cite{Fong:2010qh}, and relevant cross sections are given in 
Ref.~\cite{Plumacher:1997ru}.

Figure~\ref{Fig:M1dep_SUSY+flavor} shows the $M_{1}$ dependence of 
the lepton asymmetry $|Y_{B-L}|$ and the partial asymmetry of each lepton 
flavor $|Y_{\Delta_{i}}|$. From the Fig.~\ref{Fig:M1dep_SUSY+flavor}, 
the enhancement factor $16 \leq M_{1}/M_{1}^{0} \leq 17$ can 
yield the observed baryon number. 
The $M_{1}$ dependence of the asymmetries are 
described by the washout effects and production efficiencies of the asymmetries, 
that is basically the same  
as in the non-SUSY+flavor case. 
%%%
%%%
The magnitude of washout of each lepton flavor is parametrized by 
$K_{i}^\text{SUSY}$, which is SUSY extension of $K_{i}^\text{SM}$ 
[Eqs.~\eqref{Eq:Ke_SM} - \eqref{Eq:Ktau_SM}]: 
%%%
%%%
\begin{equation}
\begin{split}
	K_{e}^\text{SUSY} 
	= 
	\frac{\Gamma^\text{SUSY}_{N_{1} \to l_{e}H} (T=0)}
	{H (T=M_{1})} 
	\simeq 
	1.9 \left( 
	\frac{5.7 \times 10^{7} \, \text{GeV}}
	{M_{1}} 
	\right), 
	\label{Eq:Ke_SUSY}
\end{split}      
\end{equation}
%%%
%%%
\begin{equation}
\begin{split}
	K_{\mu}^\text{SUSY} 
	= 
	\frac{\Gamma^\text{SUSY}_{N_{1} \to l_{\mu}H} (T=0)}
	{H (T=M_{1})} 
	\simeq 
	8.8 \left( 
	\frac{5.7 \times 10^{7} \, \text{GeV}}
	{M_{1}} 
	\right), 
	\label{Eq:Kmu_SUSY}
\end{split}      
\end{equation}
%%%
%%%
\begin{equation}
\begin{split}
	K_{\tau}^\text{SUSY} 
	= 
	\frac{\Gamma^\text{SUSY}_{N_{1} \to l_{\tau}H} (T=0)}
	{H (T=M_{1})} 
	\simeq 
	40 \left( 
	\frac{5.7 \times 10^{7} \, \text{GeV}}
	{M_{1}} 
	\right). 
	\label{Eq:Ktau_SUSY}
\end{split}      
\end{equation}
%%%
As in the non-flavor case, these $K$ factors become $\sqrt{2}$ times
larger than in the non-SUSY case. These 
corrections make the washout of each asymmetry stronger compared with 
non-SUSY case and the SUSY effects become weak especially in the 
strong washout regime.
%%%
%%%
However, even if $K^{\rm SM}>1$, some of the $K_i^{\rm SUSY}$ can be
smaller than one, and therefore, the washout effect for the flavor $i$
is negligible. Then the supersymmetric contribution become
sizable. 
Consequently, the lepton asymmetry generation is sufficiently boosted compared 
with the case of the non-flavor
especially when $K^{\rm SM}>1$. 
%%%

Figure~\ref{Fig:evo_16M0_SUSY+flavor} shows the evolutions of total 
lepton asymmetry $|Y_{B-L}|$ and partial asymmetries of each lepton 
flavor $|Y_{\Delta_{i}}|$ for $M_{1}/M_{1}^{0} = 16$. 
%%%
%%%
To understand the importance of the flavor effects in SUSY calculation, 
we make 
Figure~\ref{Fig:flavor_ratio} in which the ratios of SUSY lepton asymmetry
to non-SUSY lepton asymmetry are plotted.  
First of all, the figure shows that SUSY enhancement factor is larger 
in the weak washout regime than in the strong washout regime as explained
in the previous subsection. Next, the figure shows that the SUSY enhancement 
factor with the flavor effect is larger than without the flavor 
effect.  Especially, it is important that even in the strong washout 
regime, the SUSY enhancement factor 
$|Y_{B-L}|_{\rm SUSY}^{\rm f}/|Y_{B-L}|_{\rm SM}^{\rm f}$ become sizable due to
the enhancements of the muon and electron asymmetry, because
$|Y_{B-L}|_{\rm SM}^{\rm f}$ has already been fairly larger than 
$|Y_{B-L}|_{\rm SM}^{\rm non-f}$ in the strong washout regime.

We could confirm the successful baryon asymmetry in the SUSY+flavor case, namely, 
in a realistic situation of the $E_{6} \times U(1)_{A}$ GUT model. 
It is sufficient to take the lightest RH neutrino mass 
$M_1\sim 16\times M_1^0\sim 9\times 10^8$ GeV for the 
observed baryon asymmetry. It is important for this calculation that
all components of neutrino Yukawa matrix are determined by the symmety
in the $E_6\times U(1)_A$ GUT and we can integrate the flavor effects on
the lepton asymmetry.

%%%%%%%%%%%%%%%%%%%%%%%%%%%%%%%%%%%%%%%%%%% 
\section{Summary and Discussion}  \label{Sec:summ} %%%%%%%%%%%%%%%%%
%%%%%%%%%%%%%%%%%%%%%%%%%%%%%%%%%%%%%%%%%%%

We have investigated the thermal leptogenesis in the $E_{6} \times U(1)_{A}$ 
GUT model in which realistic quark and lepton masses and mixings are 
obtained and the doublet-triplet splitting problem is solved with natural 
assumption that all interactions including higher dimensional interactions
are introduced with $O(1)$ coefficients.
 Each of three fundamental representations {\bf 27} includes two 
SM singlet fields, $S$({\bf 1'}) and $N_{R}^{c}$({\bf 1}), and these singlet 
fields play a role of RH neutrinos $N_{\alpha} \, (\alpha = 1, 2, ..., 6)$. 
One of the aim of this work is to show a sufficient lepton asymmetry is 
generated by the CP asymmetric decays of the lightest RH neutrino.
In the model, Majorana masses of the RH neutrinos $M_{\alpha}$ and the 
neutrino Yukawa couplings $Y_{\alpha i}$ are determined by the $U(1)_{A}$ 
symmetry. So we can calculate the lepton asymmetry, but unfortunately
the naive calculation results in too small abundance of the lepton
asymmetry. Actually, the lightest RH neutrino mass is around 
$6\times 10^7$ GeV, which is smaller than the Ibarra's lower bound
$10^{8-9}$ GeV. Moreover, the factor $K$ and the CP asymmetry $\epsilon_{N_{1}}^{SM}$ 
are evaluated as $K\sim 40$ and $\epsilon_{N_{1}}^{\rm SM}\sim 5\times 10^{-9}$. 
Therefore, the lepton asymmetry is washed out strongly
in this scenario, and even with $K\sim 1$, the $\epsilon_{N_{1}}^{\rm SM}$ is too
small to obtain the sufficient number of lepton asymmetry.

We have shown that a key ingredient for successful leptogenesis is the enhancement 
of RH neutrino masses. The model can include a large number of higher dimensional 
interactions which, and these interaction terms 
yield additional Majorana masses after developing the VEVs of negatively
$U(1)_A$ charged fields. 
%%%
%%%
The enhancements of the RH neutrino masses enhance the CP asymmetry 
$\epsilon \propto M_{1}$ and make the decay parameter $K \propto 1/M_{1}$ 
smaller to be most efficient value $K \sim 1$. How large enhancement
factor is required for the sufficient leptogenesis? To answer this question,
we have calculated the lepton asymmetry including the effects of SUSY
and flavor in the final state of the CP asymmetric decay. The result is
that the enhancement factor $16-17$ is sufficient for the successful 
leptogenesis. About 300 mass terms are sufficient to obtain this enhancement factor, and this number looks not to be difficult to be obtained 
even in the $E_6$ GUT model.   It is important that such enhancement
of the lightest RH neutrino mass 
does not change the neutrino physics at the low energy scale. 
This is because the $E_6\times U(1)_A$ GUT has six RH neutrinos which induces the same order of the amplitude of all elements of the light neutrino mass matrix.

We have calculated the lepton asymmetry in the $E_{6} \times U(1)_{A}$ model 
in following four cases: 
(i) non-SUSY+non-flavor 
(i\hspace{-1pt}i) non-SUSY+flavor
(i\hspace{-1pt}i\hspace{-1pt}i) SUSY+non-flavor
(i\hspace{-1pt}v) SUSY+flavor. 
%%%
%%%
These calculation has shown that both the effects of lepton flavor and SUSY are important.  It is known that in the strong washout
regime lepton flavor effect becomes sizable, though SUSY contribution
is not so large. We have shown that SUSY contribution becomes
important even in the strong washout regime if lepton flavor effect is included.  The essential point is that even in the strong regime 
$K^{\rm SM}>1$, the washout effects of the muon and/or the electron
can become weak, and therefore these lepton number abundances
become sizable.

Of course, the obtained result for the enhancement factor $16-17$ ($M_1\sim 
9\times 10^8$ GeV) for the sufficient
leptogenesis is dependent on the various parameters and even
on the $O(1)$ parameters. For example, we have fixed the coefficient
of $\epsilon_{N_{1}}^{\rm SM}$ in Eq. (\ref{Eq:simple_eps_SM}) as  
two in our calculation. Since the final lepton asymmetry is proportional
to this $\epsilon_{N_{1}}^{\rm SM}$ parameter, the dependence can be read
from the Fig. \ref{Fig:M1dependence_allcase}.  When the coefficient is one,
the required enhancement factor becomes around 25 
($M_1\sim 1.4\times 10^9$ GeV), and therefore,
$O(600)$ mass terms are needed. When the coefficient is four, it becomes
around 10 ($M_1\sim 6\times 10^8$ GeV), which is required $O(100)$ mass terms. 
Therefore, we will not predict the mass of the lightest RH neutrino for
sufficient leptogenesis, because it depends on the various parameters.
An important thing is that the $E_6\times U(1)_A$ GUT can explain the
baryon asymmetry in the universe.

It is not plausible to produce the sufficient lepton number in 
the $SO(10)\times U(1)_A$ GUT\cite{SO10}
 by the enhancement of the lightest
RH neutrino. Since the number of the RH neutrinos is three in $SO(10)$
model, the other two neutrinos must have the masses expected by the
symmetry. However, the difference between the $U(1)_A$ charges of
the lightest RH neutrino and the second lightest RH neutrino mass terms
is just two,
and therefore, it is not reasonable to expect that the lightest RH neutrino
has $O(10)$ times larger mass than the second lightest RH neutrino. However, since the $E_6\times U(1)_A$ has
six RH neutrinos and the difference between the $U(1)_A$ charges of
the lightest RH neutrino and the forth lightest RH neutrino mass terms
is six, it is plausible  that the lightest RH neutrino has $O(100)$ times 
larger number of  mass terms than the forth lightest RH neutrino. 
The observed baryon asymmetry in our universe may be an indirect 
signature of $E_6$ GUT. 

%%%%%%%%%%%%%%%%%%%%%%%%%%%%%%%%%%%%%%%%%%%%
\section*{Acknowledgments}   %%%%%%%%%%%%%%%%%%%%%%%%%%%%%
%%%%%%%%%%%%%%%%%%%%%%%%%%%%%%%%%%%%%%%%%%%%

This  work was supported in part by the Grant-in-Aid for the Ministry of Education,
Culture, Sports, Science, and Technology, Government of Japan,
No. 15K05048 (N.M.) and No. 25003345 (M.Y.).

%%%%%%%%%%%%%%%%%%%%%%%%%%%%%%%%%%%%%%%%%%%%%%
 %%%%%%%%%%%%%%%%%%%%%%%%%%%%%%%%%%%%%
%%%%%%%%%%%%%%%%%%%%%%%%%%%%%%%%%%%%%%%%%%%%%%%%
\end{document}